\documentclass[sigconf]{acmart}

\AtBeginDocument{%
  }

\usepackage{graphicx}
\usepackage{xcolor}
\usepackage{hyperref}
\usepackage{listings}
\usepackage[frozencache]{minted}
\usepackage{multirow}
\usepackage{listings-rust}
\usepackage{listings-riscv}
\usepackage{wasysym}
\usepackage{pifont}
\usepackage{tikz}
\usepackage{framed}
\usepackage{diagbox}
\usepackage{multicol}
\usepackage{caption}
\usepackage{subcaption}
\usepackage{paralist}
\usepackage{macrodefs}

\ccsdesc[500]{Security and privacy~Software and application security}
\ccsdesc[300]{Computer systems organization~Architectures}

\keywords{Rust; Memory safety; Capability-based security}

\copyrightyear{2025}
\acmYear{2025}
\setcopyright{cc}
\setcctype{by-sa}
\acmConference[CCS '25]{Proceedings of the 2025 ACM SIGSAC Conference on Computer and Communications Security}{October 13--17, 2025}{Taipei, Taiwan}
\acmBooktitle{Proceedings of the 2025 ACM SIGSAC Conference on Computer and Communications Security (CCS '25), October 13--17, 2025, Taipei, Taiwan}\acmDOI{10.1145/3719027.3744861}
\acmISBN{979-8-4007-1525-9/2025/10}

\begin{document}

\begin{abstract}

The Rust programming language enforces
three basic \emph{Rust principles},
namely
ownership, borrowing, and AXM
(Aliasing Xor Mutability) to prevent security bugs
such as memory safety violations and data races.
However, Rust projects often have \emph{mixed code}, i.e., code that also uses unsafe Rust, FFI (Foreign Function Interfaces),
and inline assembly for low-level control.
The Rust compiler is unable to statically enforce
Rust principles in mixed Rust code which
can lead to many security vulnerabilities.
In this paper, we propose \codename{}, a security enforcement mechanism 
that can run at the level of machine code and detect Rust principle violations at run-time in mixed code.
\codename{} is kept simple enough to be implemented into recent capability-based hardware abstractions that provide low-cost spatial memory safety.
\codename{} introduces a novel \emph{revoke-on-use}
abstraction for capability-based designs, wherein accessing a
memory object via a capability \emph{implicitly}
invalidates certain other capabilities pointing to it, 
thereby also providing temporal memory safety automatically,
without requiring software to explicitly specify such invalidation.
Thus, \codename{} is the first mechanism capable of providing cross-language enforcement of Rust principles.
We implemented a prototype of \codename{} on
QEMU.
Evaluation results show that
\codename{} is highly compatible with existing Rust
code (passing $99.7\%$ of the built-in test cases
of the 100 most popular crates)
and flags Rust principle violations in
real-world Rust projects that use FFI or inline assembly.
We discovered 8 previously unknown bugs
in such crates in our experiments.

\end{abstract}

\title{Securing Mixed Rust with Hardware Capabilities}

\author{Jason Zhijingcheng Yu}
\orcid{0000-0001-6013-157X}
\authornote{Both authors contributed equally to this research.}
\affiliation{%
  \institution{National University of Singapore}
  \country{Singapore}
}
\email{yu.zhi@comp.nus.edu.sg}

\author{Fangqi Han}
\authornotemark[1]
\affiliation{%
  \institution{National University of Singapore}
  \country{Singapore}
}
\email{fangqi_han@u.nus.edu}

\author{Kaustab Choudhury}
\affiliation{%
  \institution{National University of Singapore}
  \country{Singapore}
}
\email{kaustab.c@gmail.com}

\author{Trevor E. Carlson}
\orcid{0000-0001-8742-134X}
\affiliation{%
  \institution{National University of Singapore}
  \country{Singapore}
}
\email{tcarlson@comp.nus.edu.sg}

\author{Prateek Saxena}
\orcid{0000-0002-1875-8675}
\affiliation{%
  \institution{National University of Singapore}
  \country{Singapore}
}
\email{prateeks@comp.nus.edu.sg}

\maketitle

\ifabstractonly
\else

\section{Introduction}

The last decade has seen a rapid growth in popularity of the Rust programming language, driven by its promise of memory safety.
High-profile projects such as Linux~\cite{jonathancorbetFirstLookRust2022},
Chromium~\cite{SupportingUseRust2023}, and Windows~\cite{claburnMicrosoftRewritingCore2023} have started to adopt Rust in their development.
The security of Rust programs, however, comes with caveats.

Rust statically enforces a set of strict rules, which we call the \emph{Rust principles},
that are designed to provide memory safety as well as thread safety in Rust code.
However, these static checks restrict program expressiveness,
so Rust programs often resort to implementing some functionality in \emph{unsafe} Rust.
Unlike safe Rust, in explicitly marked unsafe Rust code blocks,
the compiler only \emph{assumes,} instead of enforcing,
the Rust principles.
This reopens the floodgates to unsafety as 
unintentionally violating Rust principles in unsafe Rust is easy.
As of this writing, for example,
RustSec~\cite{RustSecRustSecAdvisory}, a Rust security advisory database,
has catalogued over 180 memory corruption
vulnerabilities.

Moreover, Rust is 
a systems programming language where efficiency and compatibility are crucial.
Rust programs often have to interact with software written in other languages.
For example, unsafe Rust code often invokes external libraries written in C/C++ or contains
inline assembly for low-level or performance-critical operations.
Prior works report that 44.6\% of
unsafe functions are linked to foreign items~\cite{astrauskasHowProgrammersUse2020}.
Therefore, enforcing Rust principles in Rust code alone is insufficient.

In this work, our goal is to create a system that can detect run-time violations of Rust principles in \emph{mixed Rust} code,
which consists of both safe and unsafe code blocks,
and potentially also includes use of FFI and inline assembly.
Specifically, we want to extend enforcement of Rust principles \emph{across language boundaries}.
Such a system can be used to run tests on Rust programs to catch violations where they occur, much in the spirit of memory safety checkers commonly used in fuzzing or pre-deployment testing.

Dealing with cross-language enforcement is the key challenge.
It is tempting to consider implementing run-time detectors for Rust principle violations 
in all language compilers of interest as a conceptual solution.
However, this is a cumbersome approach:
it needs continuous update as language compilers evolve.
Miri, a tool for run-time detection of Rust undefined behaviours~\cite{RustlangMiri2024},
is capable of enforcing dynamic versions of the Rust principles~\cite{villaniTreeBorrows2023,jungStackedBorrowsAliasing2020} and detecting their violations
accurately.
However, Miri relies on MIR (Mid-level IR), a Rust-specific high-level intermediate representation,
which contains sufficient Rust semantics such as information about reference borrowing and ownership transfer.
It
is thus unable to enforce these principles in any language other than
pure Rust.
Consequently, it also does not work for mixed Rust programs.

Another option is to consider a weaker goal, namely that of employing a memory safety checker as a proxy for catching violations of Rust principles. For example, one can use a full memory safety checker like SoftBound+CETS~\cite{cets,nagarakatteSoftBoundHighlyCompatible2009}
or a partial safety checker like AddressSanitizer~\cite{serebryanyAddressSanitizerFastAddress2012}, to detect run-time memory errors.
However, a detector for memory safety violation is \emph{not} the same as a checker for Rust principles. We want to detect the violation of Rust principles when and where it occurs, not its after-effect that might manifest in a memory error somewhere later in the program execution. 
Violating Rust principles causes undefined behaviour in normal testing, which may or may not lead to
memory errors. 
Such violations can also
result in security-related logic errors,
e.g., reading the wrong state in security-sensitive data in privileged checks, that are not necessarily a memory safety violation.
In our experiments, Rust crate maintainers consider several of our reported Rust violations as serious enough to fix.
\newtext{We discuss the importance of detecting Rust principle violations as opposed to
only memory errors in Section~\ref{subsec:principles}.
}

\nparagraph{Our work}
We answer the above research question by presenting \codename{},
a capability-based
hardware abstraction that captures Rust principles at the \emph{machine code} level
in a language-agnostic manner, thus achieving cross-language Rust principle checking.
Our key observation is that we can abstract out the core principle from
the aliasing models underlying the Rust principles~\cite{villaniTreeBorrows2023,jungStackedBorrowsAliasing2020} into a mechanism that we call \emph{revoke-on-use}.
In the revoke-on-use mechanism, an existing pointer can derive new ones through borrowing,
similar to reference borrowing in Rust.
Upon each use of a pointer for accessing memory,
the revoke-on-use mechanism
requires that the architecture implicitly revoke (i.e., invalidate)
the pointers that conflict with the access,
which is defined based on both the borrowing relationship and the permissions of the capabilities.

We show that the revoke-on-use mechanism is well-suited to be implemented as a hardware capability for memory access. Specifically, we show how to
extend a prior capability-based design that supports flexible memory delegation and revocation~\cite{yuCapstoneCapabilitybasedFoundation2023a}.

We have implemented a prototype of the \codename{} architecture
on QEMU~\cite{bellardQEMUFastPortable2005} based on 
the 64-bit RISC-V architecture~\cite{watermanRISCVInstructionSeta}.
To enable bug detection in Rust programs, 
\codename{} requires instrumentation
of the Rust code to perform borrowing based on the pointers or references
involved in each operation.
We modified the default heap memory allocator
of Rust as well as the \texttt{rustc} compiler to
inject such \codename{}-specific borrow instructions.
All external code requires no special instrumentation.

\nparagraph{Artefacts}
We have publicly released the code and the data used in this work on Zenodo~\cite{yuArtifactsSecuringMixed2025}.

\nparagraph{Experimental evaluation} 
To evaluate the practicality of \codename{}, we focus on its effectiveness in bug finding as compared to Miri~\cite{RustlangMiri2024},
the official tool for detecting undefined behaviours in Rust programs.
We perform tests with the $100$ most popular
crates on crates.io~\cite{CratesioRustPackage}.
They include both safe Rust and mixed Rust code.
First, \codename{} is fairly compatible with real-world Rust code. It is able to pass
$99.7\%$ of the default test cases that come with the crates.
Second, by applying \codename{} to additional crates that Miri
cannot handle due to unsupported FFI or inline assembly,
we were able to discover $8$ previously unknown bugs
that violate Rust principles with their \emph{normal built-in tests}.
We reported them to the corresponding maintainers, and several of them have 
since been patched.
The results show that despite being
an abstraction of the Rust principles suitable for hardware-level implementations,
the revoke-on-use mechanism 
is effective in finding bugs in mixed Rust code already with our QEMU implementation.
Lastly, we compare \codename{} with AddressSanitizer~\cite{serebryanyAddressSanitizerFastAddress2012}, a popular and well-maintained partial memory safety
checker,
and ThreadSanitizer~\cite{serebryanyThreadSanitizerDataRace2009}, a thread-safety sanitizer.
In our experiments, \codename{} catches strictly more bugs than those alternative detectors,
highlighting the advantage of checking for Rust principles over the other respective notions.

Our QEMU-based \codename{} implementation also achieves over twofold 
run-time performance improvement over Miri on average.
We expect better performance in a hardware implementation,
though we leave that as future work.

\nparagraph{Contributions}
We propose the revoke-on-use mechanism, a novel way to enforce Rust principles across languages at the level of machine code. 
It is implementable on recent hardware capability designs which already provide spatial safety, to further provide temporal safety and maintain Rust principles.
Our QEMU-based \codename{} prototype yields a readily usable tool for detecting Rust principle violations in mixed Rust code.

\section{Overview}
\label{sec:background}

Rust statically enforces three principles on pointer use
that prevent security bugs including memory safety violations, data races,
and unsafe aliasing.
We review these key Rust principles with examples of security bugs and then describe our new \codename{} abstraction.

\nparagraph{Motivating examples} 
Listing~\ref{lst:bugs} shows three snippets (using the C programming language
as an example) each with a different type of security bugs.
Case~(a) is a temporal memory safety violation, specifically a
use-after-free. The code writes to \verb|buf| after freeing the space,
corrupting any data that resides in the reclaimed space.
Case~(b) is a data race. Two threads concurrently
update the shared \verb|buf| with one of them freeing
it in the process.
The outcome depends on how the two
threads are scheduled.
If freeing is scheduled before the other thread updates \verb|buf|,
a use-after-free occurs.
Even when we ignore freeing,
the resulting value of \verb|buf[10]| is nondeterministic.
Case~(c) is an example of an unsafe aliasing leading to  corruption of security-critical data.
The function \verb|memcpy| assumes the source and the destination regions to be non-overlapping,
and when this is not the case, as in the example code, can produce corrupted data. This can lead to a wrong result at Line~5, which further causes privilege escalation with the \verb|seteuid(0)| at Line~6.

\begin{listing}[t]
\begin{minipage}[t]{0.48\linewidth}
\begin{minted}{c}
char *buf = malloc(4096);
// ...
free(buf);
// ...
memcpy(buf, inp, 4096);
\end{minted}
\begin{center}
    (a) Memory safety violation (use-after-free)
\end{center}
\vspace{0.3em}
\begin{minted}{c}
char buf1[16];
// ...
char *buf2 = buf1 + 2;
memcpy(buf2, buf1, 8);
if (memcmp(buf2, key, 8) == 0)
  seteuid(0);
\end{minted}
\begin{center}
    (c) Unsafe aliasing causing security-critical data corruption
\end{center}
\end{minipage}
\hfill
\begin{minipage}[t]{0.49\linewidth}
\begin{minted}{c}
void consumer(void *buf) {
  // ...
  ((int*)buf)[10] = 1;
  // ...
}
int main(void) {
  int *buf = malloc(sizeof(int) * 16);
  // ...
  pthread_create(&tc, NULL, &consumer, buf);
  // ...
  buf[10] = 0;
  free(buf);
}
\end{minted}
\begin{center}
    (b) Data race causing nondeterministic states and
    potential use-after-free
\end{center}
\end{minipage}
\caption{Three types of security bugs in
C code.}
\label{lst:bugs}
\end{listing}

\begin{listing}[t]
\begin{minipage}{0.48\linewidth}
\begin{minted}{rust}
let mut buf = vec![0u8; 4096];
// ...
drop(buf);
// ...
buf.copy_from_slice(&inp);
\end{minted}
\begin{center}
    (a)
\end{center}
\begin{minted}{rust}
let mut buf1 = [0u8; 16];
// ...
let buf2 = &mut buf1[2..10];
buf2.copy_from_slice(
  &buf1[..8]
);
if *buf2 == *key {
  seteuid(0);
}
\end{minted}
\begin{center}
    (c)
\end{center}
\end{minipage}
\hfill
\begin{minipage}{0.49\linewidth}
\begin{minted}{rust}
fn consumer(buf : &mut [i32]) {
  // ...
  buf[10] = 1;
  // ...
}
fn main() {
  let mut buf = vec![0i32; 16];
  // ...
  std::thread::spawn(|| {
    consumer(&mut buf[..]);
  });
  // ...
  buf[10] = 0;
}
\end{minted}
\begin{center}
    (b)
\end{center}
\end{minipage}
\caption{Rust statically rejecting those programs for violating
Rust principles, preventing the security bugs in Listing~\ref{lst:bugs}.}
\label{lst:bugs-fail-rust}
\end{listing}

\subsection{Rust Principles}

Rust
enforces three principles to prevent such bugs:
ownership, borrowing, and AXM (Aliasing Xor Mutability).
The compiler statically checks that the safe Rust code complies with them. 
The examples in Listing~\ref{lst:bugs-fail-rust}, which are Rust versions of the buggy snippets in Listing~\ref{lst:bugs},
fail to compile for violating Rust principles.
The \emph{ownership principle} in Rust associates a unique
owner reference with each memory object.
A memory object is freed and becomes unusable after its owner is dropped
(typically, when it goes out of scope), which prevents temporal safety errors. Case~(a) violates this principle: \verb|buf| is the owner of the buffer, and is dropped
at Line~3.
The compiler thus rejects the use of \verb|buf| at Line~5.

Rust relaxes the exclusive ownership principle 
through a restricted form of aliasing called borrowing.
A borrow is a new aliased reference created from an existing reference.
In effect, it suspends the reference being borrowed and temporarily
allows using the new reference in its place.
Rust enforces the \emph{borrowing principle:} a reference cannot
outlive the reference it borrows from.
In case~(a), due to the ownership principle,
the owner reference cannot access the object at Line~5.
The borrowing principle further ensures that all other aliasing references
cannot access it after Line~3.
Those two principles, along with compile-time or run-time bounds checking,
are sufficient for full memory safety.

The third key principle is \emph{AXM}, short for \emph{Aliasing Xor Mutability.} 
It enables aggressive performance optimization when combined with the first two principles,
in addition to preventing other mistakes due to data races and unsafe aliasing.
The AXM principle states that at any given point in the program, 
each object can have either a singe active (i.e., not borrowed)
mutable reference or multiple immutable references, 
but never both simultaneously.
Mutable references are those which allow mutating the object,
while immutable references are those that do not.
Case~(c) shows a violation of the AXM principle.
Lines~3--5 create a mutable reference \verb|buf[2..10]| and an immutable reference
\verb|buf[..8]|, both borrowing from \verb|buf| and aliased with each other.
Rust rejects this behaviour and thus avoids the data corruption resulting
from aliasing references in \verb|copy_from_slice|.

The AXM principle enforced across threads can also prevent unintended consequences of data races, as illustrated in Case~(b).
Lines~9--11 attempt to pass a reference that mutably borrows
\verb|buf| to a new thread.
Since we cannot statically determine when \verb|consumer|
will complete, this borrow needs to be sustained until
the end of the whole program.
However, this cannot be satisfied, because
by the borrowing and AXM principles,
the use of \verb|buf| in \verb|main| 
requires that borrowed reference to die before Line~13.
Rust prevents the data race and its potential resulting issues
by rejecting this case.

\subsection{Exploitability of Rust Principle Violations}
\label{subsec:principles}
Enforcing Rust principles statically is too restrictive for some benign scenarios.
As such, Rust allows \emph{unsafe Rust} code blocks for which
Rust principle enforcement is disabled.
For example,
unsafe Rust allows accessing memory through raw pointers (\verb|*const T| and
\verb|*mut T|), for which none of the three basic principles are enforced,
as opposed to references (\verb|&T| and \verb|&mut T|).
Programmers use unsafe Rust extensively~\cite{zhangDualNatureNecessity2023,astrauskasHowProgrammersUse2020}.
Many data structures and smart pointers in the Rust standard library
use unsafe Rust under the hood.

Beyond enabling raw pointers which are not subject
to the static checks on Rust principles,
unsafe Rust 
allows interfacing with
non-Rust software components through
FFI (foreign function interface)
calls and inline assembly.
Program behaviours behind FFI and inline assembly are
beyond the control of the Rust compiler and are also
excluded from static checks.
Such external
code written in other programming languages can also
easily violate Rust principles.
As
a programming language geared towards systems programming
with low-level control and low overhead,
the use of FFI and inline assembly is widespread in Rust.
Prior work has estimated that $44.6\%$ of unsafe functions
in Rust projects are linked to foreign items~\cite{astrauskasHowProgrammersUse2020}.

\begin{figure}[t]
    \centering
    \includegraphics[width=0.85\linewidth]{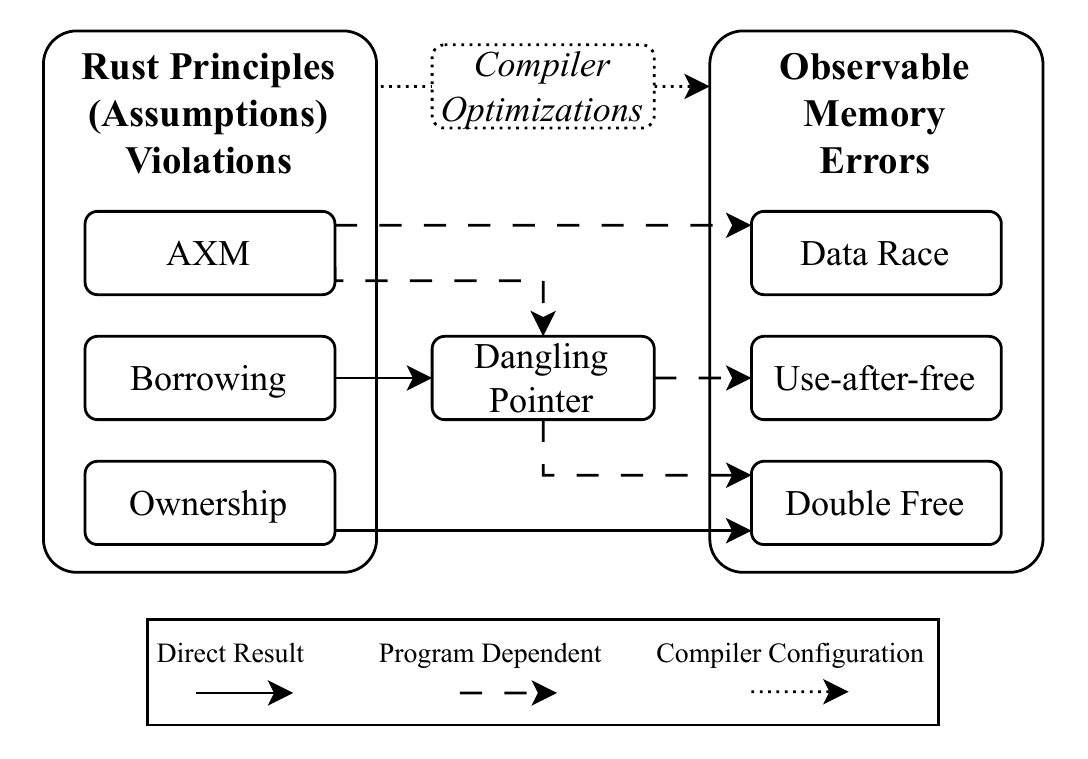}
    \caption{\newtext{A directed graph illustrating how Rust principle violations manifest into memory errors.
    }}
    \label{fig:principle-to-error}
\end{figure}

\begin{listing}[t]
\begin{minted}{rust}
fn free_xor_alias(raw: *mut u8, obj: Box<u8>) {
    assert_eq!(*obj, 0);
    unsafe { *raw = 42; }
    if *obj != 42 {
        unsafe { libc::free(raw as *mut libc::c_void); }
    }
}

fn main() {
    let mut obj = Box::new(0);
    free_xor_alias(std::ptr::addr_of_mut!(*obj.as_mut()), obj);
}
\end{minted}
\caption{\newtext{An example Rust program that incurs a logically impossible double free error after compiler optimizations.}}    
\label{lst:optimization-error}
\end{listing}

\newtext{The Rust compiler is unable to enforce the Rust principles in \emph{mixed Rust code} (i.e., safe Rust
code mixed with unsafe Rust code)
Instead,
those principles become \emph{assumptions} the Rust compiler makes.}
\newtext{Violating such assumptions can manifest
into memory errors. 
\newtext{
In particular, the Rust compiler's built-in static analyses and} 
performance optimizations assume---without checking---that unsafe Rust code preserves the Rust principles.
When the code fails to uphold these assumptions,
the resulting code can have undefined behaviour, causing
memory safety or thread safety issues as well as
incorrect execution results.
As illustrated in Figure~\ref{fig:principle-to-error},
under different compiler configurations and program implementations,
at least two paths
can lead to memory safety violations.
One path may be taken even if the program implementation does not appear exploitable, while the other path may be taken even if the compiler disables all optimizations.}
\newtext{Therefore, it is
always desirable to detect and eradicate Rust principle violations.}

Listing~\ref{lst:optimization-error} shows a Rust program where the AXM violation results in a double free error only if compiler optimizations are enabled.
The compiler assumes that, according to AXM, \texttt{raw} and \texttt{obj} in \texttt{free\_xor\_alias} are not aliased and, therefore, \texttt{obj} has been unmodified since the assertion at Line~2.
This results in the optimized program evaluating the condition at Line~4 as always \texttt{true}, causing a double free error.
The example demonstrates the unstable nature of assumption violations.
It is thus insufficient to test binaries only for particular compilations.
In practice, programmers must ensure that compiler assumptions are always satisfied.

Note that the dashed lines (i.e., program-dependent paths) in Figure~\ref{fig:principle-to-error} are also non-trivial.
Section~\ref{subsec:rust-principles-vs-sanitization} further explains how they complicate identifying bugs related to assumption violations, especially with traditional error-detecting tools.

\subsection{The \codename{} Abstraction}

In this paper, we aim to
detect Rust principle violations in mixed Rust code. 
We propose \codename{},
a new security enforcement mechanism that can be implemented at the level of machine code. \codename{} provides 
a \emph{hardware capability-based abstraction}~\cite{levyCapabilitybasedComputerSystems1984}
which detects Rust principle violations.
A capability-based abstraction replaces raw pointers
with \emph{capabilities}.
Besides the address of a memory location,
a capability carries
metadata such as memory bounds and permissions 
that restrict the access that software can
make with it.
The hardware distinguishes capabilities from
raw data and tracks their creation and use.
We choose a capability-based hardware design as the basis of our solution
for two reasons.
Firstly, such a design is language-agnostic as
hardware capabilities work at the machine code level, 
which all software in a system shares.
Inline assembly, in particular, directly exposes this level to the programmer.
Capabilities
contain extra metadata necessary to enforce
policies regarding pointer use.
In existing capability-based designs, this metadata includes
memory bounds and permissions, which are intended for catching
spatial memory safety violations directly.
\codename{} extends the capability metadata to capture
extra information needed for Rust principles.
Secondly, this design allows \codename{} to enable 
per-program full memory safety while retaining existing benefits offered by the underlying capability-based hardware, such as supporting isolation or secure data sharing which are system-wide properties.

Existing capability-based architectures
expose capability-specific instructions and
require explicit management of capabilities
by the software.
This requires
extensive changes in software across all
programming languages, including assembly code.
Our goal requires us to avoid imposing such requirements.
We propose
a novel \emph{revoke-on-use} mechanism in \codename{} to
\emph{implicitly} maintain capabilities and check
for potential violations of Rust principles.
In revoke-on-use,
when software uses a capability for a memory access,
the hardware invalidates capabilities 
whose use would conflict
with that memory access later.
The notion of \emph{conflicting} capabilities, which we will
discuss in greater detail in Sections~\ref{sec:overview} and
\ref{sec:design},
is general and simple
enough to be followed at the architectural level while
providing a useful approximation of Rust principles.

\section{\codename{}: High-level Design}
\label{sec:overview}

\codename{} is based on hardware capabilities.
In a capability-based hardware architecture, a capability
typically consists of a memory address
to access (i.e., the cursor),
permitted memory bounds, access permissions,
and other metadata such as the capability type.
Capabilities can reside in
general-purpose registers and memory, but they are distinct
from normal data and unforgeable.
The hardware maintains whether each register or
aligned memory location contains a capability or normal data,
and forbids software from converting normal data directly into
a capability.
Rather, software creates and manipulates capabilities through
dedicated instructions.
Capability-based architectures require
software to use explicitly specified
capabilities for memory accesses, 
and
perform bounds and permission checks for each
access based on the specified capability.
Such checks provide spatial memory safety.

\begin{figure}[t]
    \centering
    \includegraphics[width=0.5\linewidth]{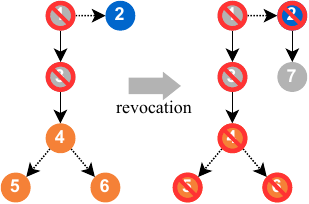}
    \caption{Overview of the revocation mechanism in \capstone{} (baseline).
    Each circle represents
    a capability, with the number indicating the order in which they are created.
    The colour indicates the capability type: linear (gray), non-linear (orange), and revocation
    (blue). Solid arrows represent capability derivations that destroy the original capabilities,
    and dashed arrows represent derivations that retain them.}
    \label{fig:capstone-overview}
\end{figure}

\subsection{Our Baseline: \capstone{}}
\label{subsec:capstone}
We choose an existing capability-based 
architecture design called
\capstone{}~\cite{yuCapstoneCapabilitybasedFoundation2023a}
as our baseline.
The reason for this choice is the additional expressiveness
of \capstone{} compared to that of other designs.
At the core of such expressiveness are linear capabilities
and revocation capabilities, two special capability types
in \capstone{}.
Linear capabilities are not duplicable.
\capstone{} guarantees
that regions pointed to by linear capabilities do not overlap with other capabilities in the system.
In tandem with linear capabilities,
revocation capabilities
enable a selective revocation mechanism, as shown in Figure~\ref{fig:capstone-overview}.
Software can pass out a linear capability as well as
delinearize it to disable the non-aliasing constraint, but
can create a corresponding revocation capability beforehand to
allow itself to
recover the same linear capability while revoking all capabilities
derived from it.
This is a default fail-close protection against accidental 
\emph{capability leaks} that affect classical capability designs~\cite{levyCapabilitybasedComputerSystems1984} 
and their modern incarnations~\cite{watsonCHERIHybridCapabilitySystem2015}.
It also provides temporal memory safety: the memory allocator
creates a revocation capability for each newly allocated object before passing the corresponding capability to the user program and
revokes with the revocation capability when the object is freed.
Beyond memory safety, 
the same capability-based mechanism in \capstone{} also
supports isolation use cases, e.g.,
memory delegation
between mutually-distrusting parties.

\begin{figure}[t]
\hfill
\begin{minipage}[b]{0.65\linewidth}
\begin{minted}[escapeinside=||]{rust}
use std::arch::asm;
fn use_p(p |\capannotate{2}| : *mut u64) {
  unsafe {
    asm!(
      "sd x0, 0({addr})",
      addr = in(reg) p |\capannotate{2}|
    );
  }
}
fn main() {
  let mut v |\capannotate{1}| = Box::new(0u64);
  let v_raw |\capannotate{2}| = &mut *v as *mut u64;
  let v_ref |\capannotate{3}| = unsafe { &mut *v_raw |\capannotate{2}| };
  use_p(v_raw |\capannotate{2}|);
  *v_ref |\capannotateinv{3}| = 42;
}
\end{minted}
\begin{center}
(a)
\end{center}
\end{minipage}
\hfill
\begin{minipage}[b]{0.3\linewidth}
\begin{center}
    \includegraphics[width=.9\linewidth]{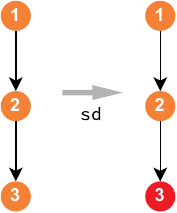} \\
    \vspace{5em}
    (b)
\end{center}
\end{minipage}
\hfill
\caption{Example of \codename{} detecting a Rust principle violation.
Circles represent capabilities. Those in red are invalid.
(a) Code with the capability associated with each
value indicated. (b) \codename{} maintains capability derivations
in trees and invalidates capabilities based on them.
}
\label{fig:big-picture-example}
\end{figure}

\nparagraph{\capstone{} falls short}
Despite the expressiveness of \capstone{},
it is not capable of achieving
our goal of detecting Rust principle violations.
We explain this with the example in Figure~\ref{fig:big-picture-example}~(a).
Line~13 dereferences a raw pointer \verb|v_raw|
to create a mutable reference \verb|v_ref|.
Line~14 then passes \verb|v_raw| as an argument
of a function call to \verb|use_p()|.
After the call returns, Line~15 writes to memory
through \verb|v_ref|.
The code violates the AXM principle because
during the lifetime of \verb|v_ref|,
the inline assembly at Lines~4-7 inside \verb|use_p()| mutates
its memory through \verb|v_raw|.
Detecting this violation requires 
1)~identifying
the derivation relationship between
\verb|v_ref| (and \verb|p| in \verb|use_p()|)
and \verb|v_raw|, 2)~identifying the write operation
through \verb|v_raw|, and 3)~invalidating
\verb|v_ref| thereafter.

In \capstone{},
to achieve the invalidation of \verb|v_ref|
when \verb|v_raw| is used, it is necessary
to create a revocation capability for
\verb|v_raw| when it is dereferenced
at Line~13, and then perform
revocation on the revocation capability
before Line~5.
Both operations require injecting extra instructions,
and looking at \verb|use_p()| alone is
not sufficient to statically
know if \verb|p| corresponds
to a revocation capability that needs revocation.
Such information may only be determined during run-time.
To handle all possibilities, therefore,
it is necessary to either
check the run-time state (e.g., the capability type)
to determine if revocation is necessary
before every memory access, or
conservatively maintain that a pointer always
has a corresponding revocation capability and
perform revocation before all memory accesses.
Both options are costly and require extensive
instrumentation to code in all languages.
The distinction between linear and non-linear capabilities in Capstone further
complicates the picture, as
linear capabilities cannot be duplicated
and non-linear capabilities cannot be used
to create revocation capabilities.

\subsection{Key Idea: Revoke-on-Use}
\label{subsec:big-picture}

\codename{} simplifies
the \capstone{} model based on a core idea which we call
\emph{revoke-on-use.}
It eschews extensive software changes while
keeping a distilled core which Rust principles and \capstone{} share.

\nparagraph{Observations}
Revoke-on-use is based on two observations.
Firstly, the act of using a capability
to access memory already carries
the intention of a program to perform any
necessary capability invalidation to allow
the access.
Instead of requiring explicit handling
of revocation capabilities, therefore,
we can simply perform revocation implicitly
when the software uses a capability to access
memory.
We thus remove the distinct revocation capability
type.
Instead, we allow the borrowing operation
to build the derivation relationship between
capabilities.
Secondly, enforcing non-duplicability of
linear capabilities is not necessary for
guaranteeing exclusive access.
A freshly borrowed capability is always unique
and the implicit revocation performed at each
memory access ensures that while the capability
is live, no access to the memory region takes place
through other capabilities.
Therefore, we also remove the distinction between
linear and non-linear capabilities.

\nparagraph{Revoke-on-use in \codename{}}
With revoke-on-use,
\codename{} provides
only one capability type.
\codename{} maintains
the derivation relationships among capabilities
as trees.
Based on the tree structures and capability metadata,
when software performs memory access using a capability,
\codename{} invalidates
other capabilities that overlap the accessed memory region
and have conflicting permissions.
In effect, revoke-on-use
ties the intention of capability revocation 
to concrete memory access.
This design choice avoids the need to statically
deduce how a pointer will be used.
It also avoids extra instructions
involving revocation capabilities.

\nparagraph{Example revisited}
We revisit the example in
Figure~\ref{fig:big-picture-example}~(a)
with revoke-on-use.
During run-time, \codename{} associates values in the program with
capabilities and maintains their derivation relationship as shown in
Figure~\ref{fig:big-picture-example}~(b).
Capability~1 is created when the software allocates
\verb|v|, and two subsequent capabilities are derived through
borrowing when the software converts the reference into a raw pointer
at Line~12
and then dereferences it at Line~13.
Capability~3 borrows from Capability~2, which in turn borrows
from Capability~1.
Capability~2 propagates along with the function call argument and into
the base address register used in the inline assembly at Line~5.
When the software performs the store operation (\verb|sd|), \codename{}
implicitly invalidates Capability~3.
This causes Capability~3 to fail the check when it is used at Line~15,
and \codename{} reports a Rust principle violation.
For \codename{} to work with this example,
the only extra information needed is for
the capability creation and borrowing operations in the Rust code.
Other parts of the code, e.g., for memory access or capability passing,
require no change.
This applies to both Rust code and code in other languages, including
inline assembly (e.g., Line~5).

\subsection{Operations on Revoke-on-Use Capabilities}
\label{subsec:revokeonuse}

We overview the operations that software can perform on a revoke-on-use capability
in \codename{}.

\nparagraph{Copy}
\codename{} allows all capabilities to be copied.
When a capability is copied, e.g., through a \verb|mv| instruction,
\codename{} treats both copies
as identical and allows them to be used interchangeably.
This avoids the problems with using linear capabilities in 
\capstone{}~\cite{yuCapstoneCapabilitybasedFoundation2023a},
as discussed in Section~\ref{subsec:capstone}.

\nparagraph{Borrow}
\codename{} allows creating a new capability through \emph{borrowing} from
an existing one.
We design such a borrowing operation to correspond to borrowing in Rust.
Unlike copying, the new capability is not treated as identical to the original even though
it points to the same memory location.
In effect, borrowing temporarily transfers exclusive access from the original capability
to the new one.
The borrowing relationship between capabilities forms a \emph{borrow tree} for each
allocation of memory object.
The root of a borrow tree is the initial capability created through allocation, and
each edge connects a capability to another that borrows from it.
A capability produced through borrowing can have its memory bounds specified as a subset of
the bounds of its parent (i.e., the capability it borrows from).
This is to support \emph{partial borrowing}, where non-overlapping subparts of a capability
are borrowed by multiple different capabilities and accessed through them independently.
Borrowing covers conversions between Rust references and raw pointers.
For example, a capability corresponding to a reference
can be created through borrowing from
a capability corresponding to a raw pointer.
This reflects dereferencing a raw pointer
to obtain a reference in Rust.
Effectively, a leaf capability corresponds to a reference or
pointer considered as currently active in Rust.

\nparagraph{Creation}
Newly allocated objects on heap or stack are associated with
capabilities.
A full capability-based system resets to an initial state with capabilities
that cover all available memory.
Memory allocators then derive capabilities for individual objects
with tighter bounds or permissions from
those initial capabilities.
This allows capabilities to confine the set of accesses a software component
can make to memory, which is important for isolation.
Since we focus on the safety of single programs rather than
isolation in this work, for simplicity,
we expose a direct $\alloc$ operation to software
which creates a new capability with specified bounds and permissions.
The new capability is the root of a new borrow tree.

\nparagraph{Revocation}
Software can perform \emph{revocation}, i.e.,
invalidate a specified capability and all capabilities derived from it (in its
borrow subtree).
This is useful when software frees a memory object.

\nparagraph{Load/Store}
Every time software performs a memory load/store, 
\codename{} checks if the capability provided for
addressing is valid.
\codename{} also invalidates conflicting capabilities
based on revoke-on-use for the allowed
memory accesses (see Section~\ref{subsec:big-picture}).

\subsection{Security Invariants}
\label{subsec:invariants}

\codename{} guarantees two simple invariants
at run-time.
Such invariants concern the properties of capabilities throughout their lifetimes (i.e., while they
are valid),
and can be understood to reflect an approximation of
Rust principles but at the machine code level.
They define when
capabilities are conflicting and require invalidation:
upon a load/store operation,
two capabilities are conflicting 
if retaining both leads to violations
of such invariants.

\nparagraph{Well-nested borrowing}
A valid capability either is the root of its borrow tree or has a parent that is also valid.
This invariant ensures that a capability cannot outlive the capability that it borrows from, which
corresponds to the borrowing principle.

\nparagraph{Exclusive access}
Consider a capability $c$. We use $D(c)$ to denote the set of
all capabilities derived from $c$ through borrowing (i.e., inside
the borrow subtree of $c$), including $c$ itself.
The exclusive access invariant states that
during the lifetime of $c$, before
any memory access to a memory location through a capability in $D(c)$,
there is
no store to an aliasing memory location from outside $D(c)$,
and before any store access to a memory location through a capability in $D(c)$,
there is no load from an aliasing memory location from outside $D(c)$.
This invariant captures the AXM principle: it guarantees
the holder of a capability \emph{exclusive access} to memory.

\subsection{Difference from Rust Aliasing Models}

The Rust compiler statically enforces Rust principles
only for safe Rust code.
Two \emph{aliasing models} of Rust,
Stacked Borrows and Tree Borrows, extend the notion
of Rust principles to run-time behaviours of unsafe Rust code.
Stacked Borrows~\cite{jungStackedBorrowsAliasing2020}
tracks references and pointers for each memory location with tags on
a stack.
Each tag has a type that carries the access permissions and aliasing requirements
associated with the pointers.
Creating a new reference or pointer through borrowing constitutes creating a
new tag and pushing it onto the stack,
and each use of a reference or pointer checks the presence of its associated tag
on the stack, its permissions, and, in some cases, pops the tags on top of it off
the stack.
Tree Borrows~\cite{villaniTreeBorrows2023} relaxes
Stacked Borrows to remove several undefined behaviours.
Unlike Stacked Borrows, it maintains references and pointers in a tree, which means it allows
but distinguishes multiple references borrowed from the same source.
For example, it allows a mutable borrow to co-exist with an aliasing immutable borrow
before the first mutation through it.

Revoke-on-use in \codename{} is generally more relaxed compared to
Stacked Borrows or Tree Borrows.
One notable difference, for example, is that
both Stacked Borrows and Tree Borrows invalidate pointers upon borrowing,
while \codename{} does so more lazily, only
when capabilities are used for accessing memory.
Another difference is that \codename{} allows all mutable references
to be reserved and treated as immutable references until
the first mutable access through them, whereas
Rust disallows this in general and only allows it in specific
scenarios to support the two-phase borrow pattern.
\codename{} also does not enforce protectors for function arguments, which
both Stacked Borrows and Tree Borrows include.
While those differences mean that \codename{} is
unable to detect all
cases of Rust principle violations, we have chosen such a
design to keep revoke-on-use capabilities simple and general
enough to be useful for other use cases of capabilities
rather than specific to Rust that can be implemented generically at machine code level.
\codename{} shows that simple adjustments to \capstone{}
allows it to detect some Rust principle violations
without losing the general usefulness of hardware
capabilities. 

\begin{table}[t]
    \centering
    \caption{\newtext{Error visibility of collected AXM violations among RustSec entries. Assessments of whether a concrete error is possible are supported by entry descriptions on RustSec.}}
    \label{tab:axm-visibility}
    \small
    \begin{tabular}{llll}
    \hline
        \textbf{RUSTSEC-} & \textbf{Type} & \textbf{Error Possible/in PoC} & \textbf{CVSS Score} \\
    \hline
2020-0023 & Vulnerability & \hspace{1.5cm} \cmark / \xmark & 9.8, Critical \\
2021-0031 & Vulnerability & \hspace{1.5cm} \cmark / \xmark & 9.8, Critical \\
2021-0114 & Vulnerability & \hspace{1.5cm} \cmark / \xmark & N/A \\
2022-0007 & Unsound & \hspace{1.5cm} \cmark / UAF & N/A \\
2022-0040 & Vulnerability & \hspace{1.5cm} \cmark / \xmark & N/A \\
\hline
    \end{tabular}
\end{table}

\subsection{Comparison with Existing Tools}

\label{subsec:rust-principles-vs-sanitization}

Existing techniques for run-time bug detection
fall in two main categories: techniques that
focus on other notions of safety rather than Rust principles,
and techniques that focus on Rust principles.
Techniques in both categories are insufficient for our
research goal.
We discuss their limitations from the perspective
of their designs below.
Sections~\ref{subsec:bug-detect} and \ref{subsec:bug-new}
show such limitations empirically.

\nparagraph{Gap between Rust principles/revoke-on-use and memory safety}
\newtext{Section~\ref{subsec:principles} has discussed how}
violations of Rust principles sometimes, but not always, cause violations
of memory safety.
The exact outcome
depends on \newtext{various factors such as} 
the specific compiler implementation
and optimizations enabled. \newtext{Furthermore, while memory risks always exist upon a violation, they sometimes require specific program implementations to generate an observable effect.
This significantly complicates program testing and undermines the language’s safety guarantees.}

\newtext{
Recall from Figure~\ref{fig:principle-to-error} that, among the three categories of assumption violations, AXM violations exhibit the most paths to various memory errors that are simultaneously the hardest to detect.
Table~\ref{tab:axm-visibility} summarizes the error visibility of our collected RustSec entries that violate AXM (Section~\ref{subsec:bug-detect} explains how they
were collected).
    Although all $5$ entries can lead to memory errors, AddressSanitizer detected only one PoC program.
The $4$ entries categorized as vulnerabilities, including $2$ with CVSS scores of $9.8$ (Critical), are undetected because their PoC implementations have not yet elevated the bug to an observable level.
Such vulnerabilities are thus undetectable by memory error detectors even on
manually crafted PoC programs.
Attempting to construct an automated approach with memory error detection tools
that cover such cases requires a more exhaustive search scheme.
}

Consequently,
existing mechanisms designed to enforce memory safety
are fundamentally incapable of reliably detecting Rust principle
violations.
These mechanisms include software-based bug detection methods
such as SoftBound+CETS~\cite{cets,nagarakatteSoftBoundHighlyCompatible2009}
and AddressSanitizer~\cite{serebryanyAddressSanitizerFastAddress2012}.
As Section~\ref{subsec:bug-detect} will show, this gap prevents them from detecting many security vulnerabilities in Rust code.

\nparagraph{Limitations of existing run-time borrow checking techniques}
\newtext{
Dedicated tools
model the root cause of the risks (i.e., assumption violation) instead of
specific symptoms.
Since such methods focus directly on the language specification,
they are capable of providing better soundness compared to
memory error detectors, covering all
legal compiler implementations and program executions.}

Miri~\cite{RustlangMiri2024} is a popular undefined behaviour
detection tool maintained by the official Rust team.
It provides run-time detection of Rust principle violations,
not just
spatial memory safety violations
and data races~\footnote{\newtext{Beyond
Rust principle violations, Miri is also capable of detecting other types
of undefined behaviours, e.g., data type confusion. We focus on
Rust principles only in this work as they are 
the most special characteristics of Rust.}}.
Miri achieves this through a run-time borrow checker
following either Stacked Borrows~\cite{jungStackedBorrowsAliasing2020} or Tree Borrows~\cite{villaniTreeBorrows2023}, selectable
through a flag.
However, to enforce an aliasing model, Miri 
requires run-time object-level information (i.e.,
which object a reference or pointer points to),
Rust-specific type information such as whether
a reference is mutable, as well as pointer provenance
(i.e., how each pointer was created).
Such information is available in MIR,
a high-level IR which Miri interprets, but
missing at the machine code level.
Therefore,
Miri is unable to support FFI or inline
assembly.
When it encounters an FFI call or
an inline assembly block, it reports an unsupported
operation and aborts the run.
We will show in Section~\ref{subsec:bug-new} that this 
limitation causes Miri to miss Rust principle violations
in real-world code.

\section{\codename{}: Design Details}
\label{sec:design}

\subsection{Capabilities}

\begin{figure}[t]
    \centering
    \small
    \begin{align*}
     c & ::= (a, a, p, b) & \text{capabilities} \\
     p & ::= \text{RW} \mid \text{RO} \mid \text{NA} & \text{permissions} \\
     b & ::= \bot \mid i & \text{parents} \\
     m & ::= \emptyset \mid m, i \mapsto c & \text{capability maps}
    \end{align*}
    $
    a \quad \text{physical addresses} \qquad i \quad \text{capabaility identifiers}
    $
    \caption{\codename{} capabilities and related constructs.}
    \label{fig:syntax}
\end{figure}

Figure~\ref{fig:syntax} defines \codename{} capabilities and related constructs.
A capability $c$ contains a pair of physical addresses encoding the
starting and ending addresses of the permitted access region.
The permission $p$ encodes the permitted access operations on the capability, which
can be read-write (RW), read-only (RO), or no-access/invalid (NA).
The parent $b$ is either null ($\bot$) or the identifier of another capability $c^\prime$ which
$c$ borrows from.
All capabilities that exist in the system are tracked in the capability map
$m$, which maps a capability identifier $i$ to the corresponding capability $m(i)$.
The parent components of all capabilities in a capability map define a collection of tree
structures (i.e., forest) over the capabilities, which we call the borrow trees.
The parent of a capability, if it is not $\bot$ (in which case
the capability is the root of a borrow tree),
then corresponds to its parent in the borrow trees.
We say a capability $c_1$ is derived from $c_2$ if $c_1$ is in the borrow subtree of
$c_2$.
\codename{} capabilities do not include the exact cursor address as in
\capstone{}.
Instead, such an address is specified by the tagged integer value itself.

The capability map defines the system state relevant to our discussion
of \codename{}.
We do not model memory and register states as they remain the same as in
existing systems, except that each address-wide location can be associated with
a capability identifier.

\subsection{Instructions}
\label{subsec:instructions}

We describe each instruction on capabilities with
a rule that derives a new capability map $m^\prime$ of the system from the previous one $m$.
When the preconditions on the rules are not satisfied, the system thus gets stuck, which
indicates that an invalid operation is encountered,
e.g., due to violating Rust principles.
Table~\ref{tab:instructions} lists the instructions and their corresponding rules, which reference auxiliary definitions
in Appendix~\ref{appx:aux-def}.
We introduce them informally with the running example
in Figure~\ref{fig:big-picture-example}.

\begin{listing}[t]
\centering
\vspace{-10pt}
\begin{multicols}{2}
\begin{lstlisting}[language={[RISCV]Assembler},moredelim={[is][\bfseries\color{orange}]{\$}{\$}},
    moredelim={[is][\bfseries\color{blue}]{\&}{\&}}]
 use_p:
  &sd&      x0, 0(a0)
  ret
 main:
  # ...
  li      a1, 8
  mv      a0, a1
  call    exchange_malloc
  $create$  a0, a0, 8
  &sd&      x0, 0(a0)
  &sd&      a0, 24(sp)
  $borroww$ a0, a0
  &sd&      a0, 16(sp)
  $borroww$ a0, a0
  &sd&      a0, 8(sp)
  call    use_p
  &ld&      a1, 8(sp)
  li      a0, 42
  &sd&      a0, 0(a1)
  # ...
\end{lstlisting}
\end{multicols}
\vspace{-20pt}
\caption{Relevant 
64-bit RISC-V binary code corresponding to the example in 
Figure~\ref{fig:big-picture-example} instrumented with
\codename{} instructions. The code is simplified
to serve an illustrative purpose
and does not exactly match our implementation.
Instructions in orange are injected,
and instructions in blue are extended with
\codename{} load/store operations.
}
\label{lst:instrumented-code}
\end{listing}

Listing~\ref{lst:instrumented-code} shows the
binary code (in 64-bit RISC-V) corresponding to the example
instrumented with \codename{} instructions.
The \verb|main()| function stores
\verb|v|, \verb|v_raw|, and \verb|v_ref|
at offsets 24, 16, and 8 to the stack pointer \verb|sp|, respectively.
A $\alloc$ instruction (Line~9) injected immediately after the
heap memory allocation (Line~8) creates a new capability
with bounds that match the address range of the allocation.
The new capability 
has a permission
of read-write and is the root of its borrow tree (i.e.,
it has $\bot$ as its parent).
Subsequent injected $\borrowmut$ instructions at Lines~12 and 14
produce capabilities that correspond to \verb|v_raw| and \verb|v_ref| respectively.
The two capabilities cover the same range, but each has
the earlier one as the parent, signifying that it borrows from the latter.
\codename{} extends the behaviours of
the load/store instructions (\verb|ld| and \verb|sd|).

At Line~2 in \verb|use_p()|, 
the store instruction (from the inline assembly)
invalidates capabilities in the subtree
of the capability for \verb|a0| at this point, which is the one
created at Line~12.
This invalidates the capability for \verb|v_ref| (created
at Line~14).
Invalidation is represented in Table~\ref{tab:instructions}
in the $\revoke(m, D_m(S))$ auxiliary function,
which returns a capability map that is the same as $m$
except that the permissions of all capabilities
identified by identifiers in the set $D_m(S)$ are $\text{NA}$
in it.
The set $D_m(S)$ contains each identifier in $S$
along with identifiers of all capabilities in its subtree.
$S$ contains all capabilities overlapping with the accessed
address $a$ (denoted as the set $O_m(a)$) which are not ancestors
of the capability identified by $i$ (denoted as the set $A_m(i)$.
Putting it together, the $\storeinsn$ instruction invalidates capabilities
in the subtree of any capability that overlaps with
the accessed address and is not an ancestor of the capability used
(including itself) in the borrow tree.
The store instruction at Line~19 accesses memory using \verb|a1| which
carries the capability of \verb|v_ref|.
The instruction checks this capability,
finds it invalid (i.e., with
permission $\text{NA}$), and hence reports a Rust principle violation.

\begin{table}[t]
    \centering
    \caption{\codename{} instruction semantics.}
    \label{tab:instructions}
    \small
    \begin{tabular}{p{1.4cm}|p{6.2cm}}
    \hline
    \textbf{Instruction} & \textbf{Rule} \\
    \hline
    $\alloc$ \newline $(a_l, a_r)$ &
    $a_l < a_r$ \quad $i \not\in \text{dom}(m)$ \newline
    $m^\prime = m[i \mapsto (a_l, a_r, \text{RW}, \bot)]$
    \\
    \hline
    $\borrowimm$ \newline $(i, a^\prime_l, a^\prime_r)$ & 
    $m(i) = (a_l, a_r, p, b)$ \quad
    $p \neq \text{NA}$ \quad
    $i^\prime \not\in \text{dom}(m)$ \newline
    $a_l \leq a^\prime_l < a^\prime_r \leq a_r$ \quad
    $m^\prime = m[i^\prime \mapsto (a^\prime_l, a^\prime_r, \text{RO}, i)]$
    \\ \hline
    $\borrowmut$ \newline $(i, a^\prime_l, a^\prime_r)$ & 
    $m(i) = (a_l, a_r, p, b)$ \quad
    $p = \text{RW}$ \quad
    $i^\prime \not\in \text{dom}(m)$ \newline
    $a_l \leq a^\prime_l < a^\prime_r \leq a_r$ \quad
    $m^\prime = m[i^\prime \mapsto (a^\prime_l, a^\prime_r, \text{RW}, i)]$ \\
    \hline
    $\dropinsn$ \newline $(i)$ & 
    $m(i) = (a_l, a_r, p, b)$ \quad
    $m^\prime = \revoke(m, S)$ \newline
    $p \neq \text{NA}$ \quad
    $S = \derived_m(R_m(i)) \cup \{R_m(i)\}$
    \\ \hline
    $\loadinsn$ \newline $(i, a)$ &
    $m(i) = (a_l, a_r, p, b)$ \quad
    $a_l \leq a < a_r$ \quad
    $p \neq \text{NA}$ \newline
    $m^\prime = \revoker \left(m, \derived_m(S)\right)$ \quad
    $S = \overlap_m(a) \cap \overline{A_m(i)}$
    \\ \hline
    $\storeinsn$ \newline $(i, a)$ &
    $m(i) = (a_l, a_r, p, b)$ \quad
    $a_l \leq a < a_r$ \quad
    $p = \text{RW}$ \newline
    $m^\prime = \revoke \left(m, \derived_m(S)\right)$ \quad
    $S = \overlap_m(a) \cap \overline{A_m(i)}$
    \\ \hline
    \end{tabular}
\end{table}

\subsection{Safety}
\label{subsec:safetyproof}

As with other capability-based abstractions,
\codename{} provides spatial memory safety by
bounds-checking all memory accesses with bounds encoded in
capabilities.
Additionally, we present
proof sketches which show
that the \codename{} design maintains the invariants
described in Section~\ref{subsec:invariants}.

\nparagraph{Well-nested borrowing}
Assume that at some time point,
a valid capability with
identity $i_1$ has an invalid parent $i_2$.
From Table~\ref{tab:instructions}, the only source of
invalid capabilities (i.e., those with the $\text{NA}$ permission)
is the $\revoke$ function in $\dropinsn$ and $\storeinsn$ instructions.
Since the capability identified by
$i_2$ is invalid, there must exist $m, m^\prime, S$,
such that $m^\prime = \revoke (m, S), i_2 \in S, m(i_1) = (-, -, -, i_2)$,
and $m, m^\prime$ are two consecutive capability map states
during run-time.
Observe that in both $\dropinsn$ and $\storeinsn$, $S$ is of the form
$S = D_m(S^\prime)$.
By induction on the definition of $D_m(S^\prime)$,
$i_1 \in D_m(S^\prime) = S$, hence
$m^\prime(i_1) = (-, -, \text{NA}, -)$, which
contradicts the assumption that the capability identified by
$i_1$ is valid.
Therefore, lifetimes of capabilities in borrow trees
are always well-nested.

\nparagraph{Exclusive access}
Assume that at a time point the capability map state is $m_0$ and
$m_0(i_1) = (a_l, a_r, -, -)$,
and $i_2 \not\in D_{m_0}(i_1)$ is the identifier of a capability in
the subtree of $i_1$.
Without loss of generality, suppose
$m_0 \longrightarrow^{\storeinsn(i_2, a)} m_1 \rightsquigarrow m_2, a_l \leq a < a_r$, and
$m_2 \longrightarrow^{\loadinsn(i_1, a^\prime)} m_3$ or
$m_2 \longrightarrow^{\storeinsn(i_1, a^\prime)} m_3$.
First, it is obvious from definitions that $i_2 \in D_{m_0}(i_1)$
if and only if $i_1 \in A_{m_0}(i_2)$.
Therefore, $i_1 \not\in A_{m_0}(i_2)$.
Since $a_l \leq a < a_r$, $i_1 \in O_{m_0}(a)$.
By semantics of $\storeinsn(i_2, a)$, $m_1(i_1) = (a_l, a_r, \text{NA}, -)$.
Observe that no instruction changes the permission of a capability
from $\text{NA}$ to $\text{RW}$ or $\text{RO}$.
Therefore, $m_2(i_1) = (-, -, \text{NA}, -)$.
With $m_2$, the conditions specified 
in semantics of neither $\loadinsn(i_1, a^\prime)$ nor
$\storeinsn(i_1, a^\prime)$ are satisfied.
This contradicts the assumption that
$m_2 \longrightarrow^{\loadinsn(i_1, a^\prime)} m_3$ or
$m_2 \longrightarrow^{\storeinsn(i_1, a^\prime)} m_3$.
Thus, before a capability is last used for memory access,
it is guaranteed that no mutation to the memory region it covers
is made through capabilities other than those in its subtree.
Similarly, we can conclude that before the last store access
with a capability, no load to its memory region is made
through capabilities other than those in its subtree.

\section{\codename{}: Implementation}
\label{sec:impl}

We have implemented a \newtext{software-based}
prototype of \codename{} as a readily usable \newtext{debugging} tool
for detecting Rust principle violations in mixed Rust code.
The implementation consists of two components:
1)~a capability-based architecture, which adds support for encoding
and maintaining \codename{} revoke-on-use capabilities and related instructions using QEMU~\cite{bellardQEMUFastPortable2005},
and
2)~\codename{}-specific instrumentation to Rust code to expose key semantics
to the hardware.
\newtext{We expect a hardware implementation to achieve better performance
and enable extending the abstraction to a full system and integration
with existing capability use cases, e.g., for software compartmentalization.}
\newtext{We plan to investigate this in future work.}

\subsection{Rust-specific Adjustments for Compatibility}
In practice, some special cases in Rust programs require adjustments
to the core \codename{} design introduced in Sections~\ref{sec:overview} and
\ref{sec:design}.
Those adjustments remove major false positives, i.e., cases with
misidentified Rust principle violations. 
We include the following Rust-specific adjustments in our implementation.

\nparagraph{Raw pointers}
Unlike references, software can use
multiple aliasing raw pointers
created from the same reference in an interleaving way.
Therefore,
for raw pointers, we enforce a different exclusiveness invariant:
while a raw pointer type capability is valid,
no store to any memory location that aliases with
the capability memory bounds takes place unless through a capability derived from its
parent through borrowing (i.e., inside the subtree of its parent in the borrow tree).
We add a \emph{type} bit to capability metadata to identify capabilities
corresponding to raw pointers.

\nparagraph{Interior mutability}
Interior mutability in Rust allows mutation even behind immutable references.
It requires such mutation to be performed behind an UnsafeCell reference.
We extend the type field further to record if a capability corresponds to an UnsafeCell,
and further relax the exclusiveness invariant to except aliasing stores from
within subtrees of UnsafeCell capabilities:
while a capability is valid (i.e., between its creation and last use)
no store to any memory location that aliases with
the capability memory bounds takes place unless through a capability derived from it
through borrowing (i.e., inside its subtree in the borrow tree), \emph{or derived from
an UnsafeCell capability it is not derived from.}

\nparagraph{Pointer provenance}
\codename{} tracks capability propagation to associate correct metadata
with pointers.
Doing this na\"ively at the machine code level
easily causes errors.
C standard library implementations, e.g., glibc,
contain widespread use of pointer arithmetic, including operations which
subtract or add pointers together.
They are purposefully done, often for performance optimizations.
For instructions that take two input register operands, e.g., \texttt{add} and
\texttt{xor}, both registers can be capabilities, in which case
it cannot immediately be determined with which capability
to associate the result.
Our implementation overcomes this problem by allowing multiple capabilities
to be associated with each value.
Instructions with two input register operands combine the lists of capabilities associated with
the inputs to produce the result.
When the software later uses the result for load or store for the first time,
our implementation resolves its provenance by keeping only the associated capability
with memory bounds matching the requested load/store address.

\subsection{Implementation on an Instruction Set}
We implement an instantiation of \codename{}
on RV64~\cite{watermanRISCVInstructionSeta}
based on user emulation mode of QEMU~\cite{bellardQEMUFastPortable2005} with the TCG
backend.

\nparagraph{Capability storage}
To maintain maximum binary-level compatibility with existing software,
we eschew changing any data in memory and instead
store capability metadata in shadow memory inaccessible to the user
program, indexed by the virtual address.
In particular,
compared with designs that store capability metadata inline in memory~\cite{watsonCHERIHybridCapabilitySystem2015,yuCapstoneCapabilitybasedFoundation2023a},
storing metadata offline has the benefit of avoiding the need to
expand the pointer width, which in turn necessitates adjustments to ABIs and
changes to all software, including external libraries the Rust program calls into.

Our implementation recycles the metadata storage for capabilities.
It maintains for each capability a reference count, corresponding to the number
of memory locations containing it.
Upon failure to allocate metadata space for a new capability due to lack
of free space, our implementation checks the reference counts of all
capabilities and scans through all registers. After combining such information,
it adds all metadata space associated with capabilities that are invalid and
appear in neither
memory nor registers to a \emph{free} linked list, from which
it allocates metadata space thereafter.

\nparagraph{Borrow tree}
Revocations in \codename{} consist of invalidating of whole subtrees of borrow trees.
We implement this with simple subtree traversals, marking each visited capability
as invalid.
Since invalidation is irreversible (an already invalid capability never becomes valid),
once invalidated, a capability stops playing any role in borrow trees.
As a performance optimization, we disconnect a subtree from the borrow tree after invalidating it, thus
avoiding paying for invalid capabilities during future
invalidations.
In this way, capabilities connected to a borrow tree are always valid.

\subsection{Rust Compiler Instrumentation}
We instrument the Rust code to inject instructions that create, free, and borrow capabilities.
Such instrumentations pass the necessary high-level semantics of Rust code to the machine code level.
External libraries require no recompilation.
The instrumentation on Rust code takes place in three different places.

\nparagraph{Rust standard library}
We modified the default heap allocator \verb|Global| in the Rust standard library to create
a capability for every heap allocation and invalidate the associated capabilities every time
software frees a heap object.
The Rust standard library was also instrumented with the
modified MIR pass described below.
\newtext{Since this prototype implementation
interposes the standard memory allocator in
Rust code but not memory allocators in foreign code,
it does not protect memory allocated in foreign code.
Note that this is not due to fundamental limitations
of the revoke-on-use model.
Interposing memory allocators for 
foreign code will enable such protection,
and
a hardware implementation of \codename{} can
protect all memory even without
interposing on memory allocators.

}

\nparagraph{MIR pass}
We added an extra MIR pass to the reference Rust compiler \texttt{rustc} to inject
borrow instructions. 
This added MIR pass
scans through the MIR code for sites of conversions between raw pointers and references,
and injects borrow instructions immediately after such conversions.
Such instructions are injected as calls to functions that include inline assembly code to
perform the borrow.
We choose to instrument at the MIR level because 
it is simpler than Rust source code
while keeping
the relevant semantics intact.

\nparagraph{LLVM frame lowering}
We modified the prologue and epilogue generation logic in the
LLVM backend to cover stack objects with capabilities.
The modified prologue creates a capability for each stack object,
while the modified epilogue performs invalidation
on the associated capabilities of stack objects one by one.

\subsection{Putting It Together}
Figure~\ref{fig:pipeline-overview} shows an overview of how the components
fit together.
Take a Rust program that links to a C library as an example.
The program is built as normal, except that the Rust code is compiled using
the modified \texttt{rustc} compiler (including the modified LLVM backend and
Rust standard library)
that injects \codename{} instructions, with
optimizations disabled to preserve as much original semantics in
the source code as possible.
The resulting binary code is then executed on the \codename{} architecture
implemented in QEMU, which then
reports any Rust principle violations encountered.

\begin{figure}[t]
    \centering
    \includegraphics[width=0.7\linewidth]{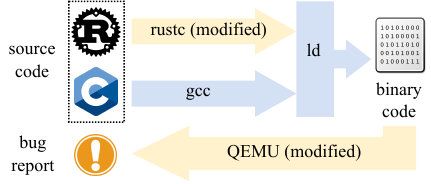}
    \caption{\codename{} in operation with
    an example program in Rust and C.
    The shade colour indicates if
    it is modified for \codename{}: blue for unmodified and orange for modified.}
    \label{fig:pipeline-overview}
\end{figure}

\section{Evaluation}
\label{sec:eval}

We answer three evaluation questions (EQs) in our experiments:

\textbf{EQ1.} Is \codename{} compatible with existing Rust code?

\textbf{EQ2.} Can \codename{} detect new bugs in mixed Rust code?

\textbf{EQ3.} What is the expected performance overhead of \codename{}?

We ran all experiments on a machine with an AMD Ryzen Threadripper 3970X 32-core processor
and 96~gigabytes of RAM, inside an Apptainer container with Ubuntu 22.04 LTS.

\begin{table}[t]
\centering
\caption{\newtext{Summary of compatibility evaluation results.
Legend: P(assed), I(gnored), U(nsupported), O(thers).}}
\label{tab:compat-stats}
\small
\begin{tabular}{|l|cccc|c|}
    \hline
     \diagbox{\textbf{Ours}}{\textbf{Miri}}
     & \textbf{P} & \textbf{I} & \textbf{U} & \textbf{O} & \textbf{Total} \\
     \hline
\textbf{P} & 4948 & 17 & 12 & 1 & 4978 \\
\textbf{OOB} & 1 & 0 & 0 & 0 & 1 \\
\textbf{Invalid cap} & 11 & 0 & 0 & 1 & 12 \\
\textbf{O} & 0 & 0 & 1 & 0 & 6 \\
\hline
\textbf{Total} & 4960 & 17 & 13 & 2 & 4992 \\
\hline
\end{tabular}
\end{table}

\subsection{Compatibility with Rust Crates}
\label{subsec:compat}

To answer EQ1, we compare the result of running existing Rust code on \codename{} against that
of running it on Miri.

\nparagraph{Benchmarks}
We collected unit tests and integration tests bundled with
the 100 most popular (by numbers of downloads) crates on crates.io~\cite{CratesioRustPackage}.
We assume that these widely-used Rust projects are representative
of idiomatic Rust code and are of relatively high quality.
From these crates, we successfully collected and built a total of 5388~test cases\footnote{
Since crate authors may configure each test case to be disabled or ignored for different
setups (e.g., target architecture, whether on
on Miri), we include only test cases that are (1) present for both setups or
(2) disabled or ignored only because of Miri.},
\newtext{from which we further exclude
those
that fail due to missing dependencies in the environment (e.g., missing software
packages) or time out on either Miri or \codename{}.
This results in a total of
4992~test cases used in our compatibility evaluation.}

\nparagraph{Miri configuration}
We first tried targeting Miri to
\riscv{}~\cite{watermanRISCVInstructionSeta}, matching \codename{}.
However, for 60 out of the 100 crates
covered,
Miri reports unsupported use of inline assembly.
Further inspection reveals
that the inline assembly is in
the \verb|core::core_arch::riscv64| and \verb|core::core_arch::riscv_shared|
modules, which provide \riscv{}-specific intrinsic implementations.
In comparison, Rust provides their counterparts in \hbox{x86-64} through
invocations of LLVM intrinsics, which Miri emulates.
We decided that Miri did not provide a mature enough support for the \riscv{} backend 
and therefore configured Miri to target \hbox{x86-64} instead
for our main experiments.

\begin{figure}[t]
    \centering
    \includegraphics[width=0.45\linewidth]{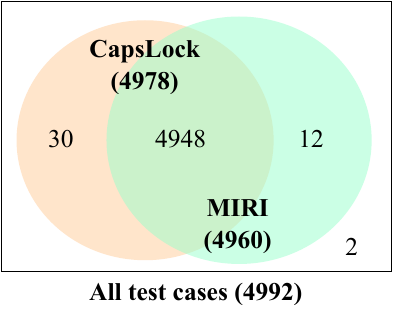}
    \caption{\newtext{A Venn diagram comparing the numbers of test cases passed on
    \codename{} and Miri. \codename{} passes $99.7\%$ of the test cases in
    the 100 most popular crates.}}
    \label{fig:venn-compat}
\end{figure}

\nparagraph{Results}
We categorize test cases based on whether they pass on \codename{} and Miri, 
as well as on the cause of failure.
Table~\ref{tab:compat-stats} provides a summary of the evaluation results and
Figure~\ref{fig:venn-compat} shows a Venn diagram of the test cases passed on
\codename{} and Miri, respectively.
A majority
of the test cases pass with no violations on \codename{} and Miri.
\codename{} passes $99.7\%$ of the test cases.

\nparagraph{Comparison with Miri}
In total, more test cases pass on \codename{} than on Miri.
\codename{} passes all test cases ignored by Miri and
$12/13$ of the tests with operations unsupported by Miri.

\newtext{
\nparagraph{Failed cases} 
A total of 12~test cases ($0.2\%$) pass on Miri but fail
on \codename{}. 
Those test cases are distributed across 5
crates: \texttt{crossbeam-utils}, \texttt{once\_cell},
\texttt{parking\_lot}, \texttt{tempfile}, and \texttt{url}.
We investigate them and identify two main
mismatching corner cases involving
interior mutability between the revoke-on-use model of
\codename{} and the proposed Rust aliasing models~\cite{villaniTreeBorrows2023,jungStackedBorrowsAliasing2020}.
Unlike revoke-on-use, proposed Rust aliasing models
allow bypassing an UnsafeCell to access memory it covers.
Since they are defined per memory location,
they also allow interleaving uses of multiple references
to memory regions covered by the same UnsafeCell, as long
as those regions are disjoint.
We did not pursue those corner cases further
as no officially accepted aliasing model exists yet.
\footnote{\newtext{%
Stacked Borrows and Tree Borrows capture slightly different requirements, with
Tree Borrows being generally more tolerant.
It is possible to adapt \codename{} to match either model more accurately, but
we avoid this for two reasons.
First, we avoid overfitting the ISA to Rust. Instead,
we focus on the goal of highlighting the generality of application-level isolation and object-level memory safety.
Second, for the Rust use case, 
we focus only on the core component of the aliasing model, i.e.,
Rust principles.
The other details of the Rust aliasing
model are still work in progress and under debate.}}}

\begin{takeaway}
\codename{} is highly compatible with existing
code, passing $99.7\%$ of 
test cases from the 100 most popular
crates on crates.io~\cite{CratesioRustPackage}.
\end{takeaway}

\subsection{Effectiveness in Security Bug Detection}
\label{subsec:bug-detect}

We evaluate the ability of \codename{} to detect known cases of
Rust principle violations. We also experimentally determine the gap
between Rust principles and two other notions:
memory and thread safety.
We find that tools designed for detecting issues for the latter are
incapable of detecting many bugs caused by Rust principle violations,
\newtext{in line with our discussion 
in Section~\ref{subsec:rust-principles-vs-sanitization}.}

\nparagraph{Benchmarks}
We collect proof-of-concept exploits (PoCs) of 15 known vulnerabilities
attributed Rust principle violations
from the RustSec Advisory Database~\cite{RustSecRustSecAdvisory}.%
\footnote{
At the time of our experiments, RustSec contained a total of 641 entries.
We used the following criteria to select the
entries to use in our experiments:
We first included those in the
\texttt{memory-corruption}, \texttt{memory-safe}, and \texttt{thread-safety} categories, and additionally those that
mentioned \texttt{memory}, \texttt{thread}, \texttt{race}, or \texttt{sound} in the keywords or the advisory main text.
For the resulting 333 entries, we then collected 162 PoCs.
We successfully built 55 of those PoCs for the 64-bit \riscv{}
Linux platform.
Among them, 15 are attributed to violations of
Rust principles and thus relevant to our experiments. 
}
We test the effectiveness of \codename{} in bug finding
on those 15 PoCs compared to Miri.
We also include results of AddressSanitizer~\cite{serebryanyAddressSanitizerFastAddress2012} and ThreadSanitizer~\cite{serebryanyThreadSanitizerDataRace2009}, two widely-deployed checkers for memory safety violation and data race detections,
to verify the gap between Rust principles and these alternative notions.

\begin{table}[t]
    \centering
    \caption{\newtext{Evaluation results on collected PoCs attributed to Rust
    principle violations. A checkmark (\cmark)
    indicates that a violation is detected.
    A double cross (\xmark\xmark) for AddressSanitizer (AS) and ThreadSanitizer (TS) indicates
    missed cases due to 
    the memory or thread safety risks not being visible during test run.
    A single cross (\xmark) indicates cases missed despite memory errors or data races visible in the run.
    }}
    \label{tab:benchmarks-cves}
    \small
    \begin{tabular}{llcccc}
    \hline
        \textbf{RUSTSEC-} & \textbf{Violation} & \textbf{\codename{}} & \textbf{AS} & \textbf{TS} & \textbf{Miri} \\
    \hline
2020-0023 & AXM & \cmark & \xmark\xmark & \xmark\xmark & \cmark \\
2021-0031 & AXM & \cmark & \xmark\xmark & \xmark\xmark & \cmark \\
2021-0114 & AXM & \cmark & \xmark\xmark & \xmark\xmark & \cmark \\
2022-0007 & AXM & \cmark & \cmark & \xmark & \cmark \\
2022-0040 & AXM & \cmark & \xmark\xmark & \xmark\xmark & \cmark \\
2019-0009 & Borrowing & \cmark & \cmark & \xmark & \cmark \\
2019-0023 & Borrowing & \cmark & \cmark & \cmark & \cmark \\
2020-0091 & Borrowing & \cmark & \cmark & \cmark & \cmark \\
2021-0130 & Borrowing & \cmark & \cmark & \xmark & \cmark \\
2022-0002 & Borrowing & \cmark & \xmark\xmark & \xmark\xmark & \cmark \\
2022-0070 & Borrowing & \cmark & \xmark & \xmark & \cmark \\
2023-0070 & Borrowing & \cmark & \cmark & \xmark & \cmark \\
2021-0039 & Ownership & \cmark & \cmark & \xmark & \cmark \\
2021-0049 & Ownership & \cmark & \cmark & \xmark & \cmark \\
2021-0053 & Ownership & \cmark & \cmark & \cmark & \cmark \\
\hline
    \end{tabular}
\end{table}

\nparagraph{Results}
\codename{} detects all 15 Rust principle violations across
all three types,
matching Miri, as shown in
Table~\ref{tab:benchmarks-cves}.

\nparagraph{Comparison with AddressSanitizer and ThreadSanitizer}
Corroborating Section~\ref{subsec:rust-principles-vs-sanitization},
tools not designed for detecting
Rust principle violations show
limited effectiveness.
AddressSanitizer and ThreadSanitizer together
only detect memory safety violations or data races in
9 of the 15 bugs that violate Rust principles.
Among the three principles, they fare most poorly
with AXM violations.
Only 1 out of 5 AXM violations (RUSTSEC-2022-0007) is detected by
AddressSanitizer as it results in a use-after-free, and none is detected
by ThreadSanitizer.
We inspect the cases they both fail to detect, and find that most
(5 out of 6) do not trigger memory safety violations or data races.

\begin{takeaway}
\codename{} is able to detect all known Rust principle violations
we tested, matching the results from Miri.
Memory safety violation and data race detection tools
fail to detect $40\%$ of such violations.
\end{takeaway}

\subsection{Improvements for Mixed Rust Code}
\label{subsec:bug-new}

Miri~\cite{RustlangMiri2024}
is the official Rust tool to catch
undefined behaviours through runtime testing in mixed Rust code.
Note that Miri does not work with code using FFI or inline assembly.
Nonetheless, it is widely used.
We examine in EQ2
whether \codename{} can detect bugs in programs that Miri does not work with,
specifically on programs with FFI
or inline assembly.

\nparagraph{Benchmarks}
We collected built-in test cases
of crates for which Miri reported ``unsupported operations''
from crates.io~\cite{CratesioRustPackage}.
Specifically,
we include the top 1000 popular crates along with 13000
randomly sampled crates deemed to potentially
use FFI.%
\footnote{We use a simple heuristic: if the source code contains
any of the strings \texttt{ffi}, \texttt{bindgen}, and \texttt{\#[link}, we
consider the crate as likely to use FFI.}
Test cases that do not build automatically through \texttt{cargo} are discarded.
In total, we obtain 86051 test cases.
Miri reports unsupported operations for 10478 of them,
which we then run on \codename{} successfully.
All crates are the latest versions available at the time of
experiments.

\begin{table*}[t]
    \centering
    \caption{Summary of previously-unknown bugs detected
    with \codename{}. For the maintainer
    response, \cmark{} indicates \emph{confirmed and fixed}, and \bell{} indicates \emph{confirmed but not 
    fixed}. 
    ``(S)'' indicates that
    Stacked Borrows~\cite{jungStackedBorrowsAliasing2020}, but not Tree Borrows~\cite{villaniTreeBorrows2023},
    considers the principle as violated, and ``(X)'' indicates that the bug can cause
    security issues under the current \texttt{rustc} implementation.}
    \label{tab:bugs-new}
    \small
    \begin{tabular}{llllc}
    \hline
    \textbf{Crate} & \textbf{Downloads (Total / 90-day)} & \textbf{Violation} & \textbf{Unsupported Operation Miri Reports} & \textbf{Response}\\ \hline
        \verb|ring 0.17.8| & 171146621 / 28216664 & AXM (S) &
        \verb|ring_core_0_17_8_OPENSSL_cpuid_setup| & \cmark \\
        \verb|generator 0.8.3| & 9520725 / 1367251 & AXM &
        \verb|getrlimit| \\
        \verb|sxd-document 0.3.2| & 877547 / 94741 & AXM &
        \verb|gnu_get_libc_version| through \verb|cargo-wix 0.3.8| & \cmark \\
        \verb|mozjpeg 0.10.10| & 395789 / 43760 & AXM &
        \verb|jpeg_std_error| & \cmark \\
        \verb|vorbis_rs 0.5.4| & 44149 / 16891 & AXM &
        \verb|vorbis_comment_init| & \cmark \\
        \verb|corgi 0.9.9| & 22508 / 6318 & AXM & \verb|mi_malloc_aligned| & \cmark \\
        \verb|cadical 0.1.14| & 15567 / 3268 & AXM & \verb|ccadical_init| & \bell \\
        \verb|libipt 0.1.4| & 7854 / 4326 & Borrowing (X) &
        \verb|pt_cpu_errata| \\
    \hline
    \end{tabular}
\end{table*}

\nparagraph{Bugs found}
\codename{} detects
8 previously unknown cases of Rust principle violations across
the FFI, which Miri is unable to handle, as summarized
in Table~\ref{tab:bugs-new}.
We have manually reviewed the relevant source code to confirm
that they are violations of Rust principles and
analyse the root causes.
We have reported these issues to the maintainers on
GitHub, which hosts the repositories of all eight crates.
When communicating the issues, we reproduced simplified
versions of the relevant code entirely in Rust to
run on Miri and
confirmed that Miri also considered them as
Rust principle violations.
In five cases, the maintainers have confirmed and fixed the bugs.
The crates show varying degrees of popularity, 
ranging from
having under ten thousand total downloads to over
twenty million monthly downloads.
This reflects the fact that violations of Rust principles do occur
and the lack of effective tools and methods
in detecting such violations in projects that involve FFI.
Most (7 out of 8) crates violate the AXM principle.
Miri fails to handle their use of FFI, including
calls to 
system APIs (e.g., \verb|getrlimit| and \verb|gnu_get_libc_version|)
and application library APIs (e.g., \verb|jpeg_std_error|).
\codename{} is able to handle both cases.
In one case (\verb|sxd-document|), \codename{} detected the
issue when running test cases of another crate that depended on
it.

It is worth noting that these bugs were found with {\em normal built-in} tests and did not require any specialized creation of proof-of-concept exploits. In order to understand whether these bugs would be caught by off-the-shelf memory safety checkers, we tried crafting specific PoCs to check if a memory safety violation can be induced from them.
We find that $1$ (\verb|libipt|) out of the $8$ bugs can lead to a use-after-free under the current \texttt{rustc} compiler implementation.
Note that despite this, the built-in test cases do not trigger any memory safety issue.
As such, tools like AddressSanitizer are unable to detect the bug.
To the best of our knowledge, the other bugs currently cause no security violations, but that is subject to changes
in \texttt{rustc} implementation and the optimization techniques used, \newtext{as
discussed in Section~\ref{subsec:principles}.
Therefore, all 8 bugs can impact security and should be addressed in a memory safe program.}
We discuss the details of three of those detected bugs
in Section~\ref{sec:casestudy}.

\begin{takeaway}
\codename{} detects $8$ previously-unknown bugs in cases
with FFI and inline assembly which Miri does not support.
\end{takeaway}

\subsection{Performance}
\label{subsec:perf}

We answer EQ3 by
measuring the time \codename{} and Miri spend running
the 4948 test cases from the compatibility experiments
(see Section~\ref{subsec:compat}) which both \codename{} and Miri pass.

\nparagraph{Results}
Figure~\ref{fig:performance-hist} shows a histogram that compares the two execution
times for each test case.
A total of $98.1\%$ of the test cases
take less time on \codename{} than on Miri.
On average, \codename{} spends $45.9\%$ as much time as
Miri on a test case.

\newtext{
\nparagraph{Discussion}
It might appear counterintuitive first how \codename{}
outperforms Miri
despite working at the lower and more detailed binary level.
We note two potential factors that contribute to \codename{}'s
performance advantage.
Firstly, Miri maintains
high-level metadata that it has access to to 
cover more types of undefined behaviours than \codename{} does.
For example, Miri also handles more general
type safety violations, e.g., using an arbitary integer
value as a boolean.
\codename{} instead focuses on violations of the three 
principles that are unique to Rust without tracking
such information.
Secondly, QEMU 
provides JIT compilation through its TCG engine~\cite{bellardQEMUFastPortable2005} 
and is optimized for high performance.
In contrast, MIRI operates entirely as an interpreter.
}

\begin{takeaway}
\codename{} achieves more than $2x$ speed-up over the present implementation of Miri on average.
\end{takeaway}

\section{Qualitative Analysis of New Bugs}
\label{sec:casestudy}

Our qualitative analysis reveals that
\codename{} can detect bugs that require various levels
of detection ability.

\begin{figure}[t]
\begin{minipage}[t]{.5\linewidth}
    \vspace{0pt}
    \centering
    \includegraphics[width=\linewidth]{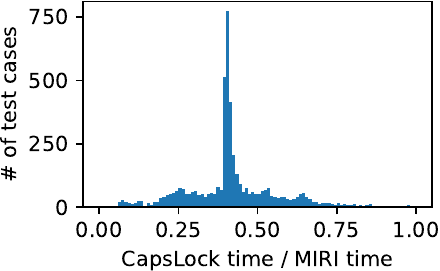}
    \vspace{-15pt}
    \captionof{figure}{Histogram of the performance of \codename{} compared to
    Miri (bin size: $0.01$).}
    \label{fig:performance-hist}
\end{minipage}
\hfill
\begin{minipage}[t]{.48\linewidth}
    \vspace{0pt}
    \centering
    \begin{minted}{rust}
    let mut r = Box::new(N::new());
    let p = &mut *r as *mut _;
    r.p = p;
    // ...
    unsafe { &mut *r.p }
    \end{minted}
    \vspace{-15pt}
    \captionof{listing}{AXM violation in \texttt{generator} (simplified).}
    \label{lst:generator}
\end{minipage}
\end{figure}

\nparagraph{The easy case: FFI not involved}
In the easiest scenario, a test case
uses FFI but the bug does not directly involve it.
To handle such cases,
it suffices to run
external code without tracking pointer use
therein.
An example is the bug found in \texttt{generator}, which
needs to allocate stack space for its managed threads.
On UNIX-like systems, the crate
invokes the \verb|getrlimit| system call
to decide the stack size limit
(\verb|RLIMIT_STACK|).
Miri does not support this system call and thus fails to
run these test cases.
The part of the crate that violates AXM (Listing~\ref{lst:generator})
is entirely in Rust and unrelated to the system call.
It handles pointers to maintain a tree data structure.
The code stores a raw pointer to \verb|r| in \verb|r.p| 
(Line~3) and then dereferences the raw pointer (Line~5).
This results in two mutable references to the same object (\verb|r|),
a violation of AXM.
Under both Stacked Borrows and Tree Borrows, \verb|p|
would become invalid
when \verb|r.p| is mutated at Line~3, making the dereferencing illegal
at Line~5.

\begin{listing}[t]
\begin{minted}{rust}
let mut started = CompressStarted { 
  compress: self, 
  dest_mgr: Box::new(DestinationMgr::new(writer, write_buffer_capacity)), 
}; 
unsafe { 
  let dest_ptr = addr_of_mut!(started.dest_mgr.as_mut().iface); 
  started.compress.cinfo.dest = dest_ptr; 
  ffi::jpeg_start_compress(&mut started.compress.cinfo, boolean::from(true)); 
}
// ...
unsafe { 
  ffi::jpeg_finish_compress(&mut started.compress.cinfo); 
} 
\end{minted}
\caption{AXM violation in \texttt{mozjpeg}.}    
\label{lst:mozjpeg}
\end{listing}

\nparagraph{Bugs caused by pointer use behind FFI}
More difficult to detect are
violations caused by pointer use in external code
after
Rust code passes the pointer out through FFI.
It is no more sufficient to merely run the external code.
Rather, it requires tracking how external code uses
the pointer.
For example,
the \verb|mozjpeg| crate provides Rust bindings for
MozJPEG, a JPEG encoder library written in C.
Miri is unable to run the test cases due to the extensive use of FFI.
Listing~\ref{lst:mozjpeg}
shows where \verb|mozjpeg| violates
AXM.
It stores a pointer to \verb|started.dest_mgr|
in \verb|started.compress.cinfo.dest| (Line~7), and later invokes
the FFI \verb|jpeg_finish_compress| with \verb|started.compress.cinfo|
borrowed and passed in as an argument.
The C function \verb|jpeg_finish_compress| then uses the raw pointer
\verb|started.compress.cinfo.dest| to mutate the content of
\verb|started.dest_mgr|.
This violates AXM as \verb|started.dest_mgr| is accessed without the owner being
borrowed.
This violation directly involves FFI.
It requires knowing that \verb|jpeg_finish_compress|
dereferences \verb|started.compress.cinfo.dest|.

\nparagraph{Bugs caused by cross-FFI pointer propagation and use}
Even more demanding is the detection of bugs that involve
pointers propagated across FFI multiple times.
The Rust code may pass pointers to data
structures to external code, which then initializes those
data structures
by storing some pointers in them.
Later,
Rust code passes in a pointer to one such data structure,
and the external code uses pointers in it to
access other data structures.
Detecting such a bug requires tracking not only how each FFI call
individually uses pointers, but also
pointer propagation in the global state
(e.g., heap memory).
For example, the \verb|vorbis_rs| crate provides Rust bindings for
a Vorbis codec library in C.
Miri fails to run the test cases due to FFI.
Listing~\ref{lst:vorbis}
shows the code related to an AXM violation.
The Rust code first allocates memory for several data structures.
It then invokes the library to perform initializations (Lines~2
and 7, top), which
connect those
data structures to one another by storing those pointers (Lines~4 and 9, middle).
The Rust code still keeps owner references to those data structures.
It then passes them when invoking more library APIs (Line~12, top).
In those library calls, those pointers are dereferenced for
mutation (Line~3, bottom).
This violates AXM similarly as in \verb|mozjpeg|---mutating an owned object without borrowing the owner---but its detection
requires more knowledge about the external code.
In particular, \verb|mozjpeg| sets up the data structure 
pointers in Rust, so
knowing if external code dereferences them is sufficient,
whereas in \verb|vorbis_rs| both initialization and use
of the data structures happen in FFI calls.
It also requires knowing where the pointers are stored during initialization.

\begin{listing}[t]
\begin{minted}{rust}
unsafe {
  libvorbis_return_value_to_result!(vorbis_analysis_init( 
    vorbis_dsp_state.as_mut_ptr(), 
    &mut *vorbis_info.vorbis_info 
  ))?; 
  let mut vorbis_dsp_state = assume_init_box(vorbis_dsp_state); 
  libvorbis_return_value_to_result!(vorbis_block_init( 
    &mut *vorbis_dsp_state, 
    vorbis_block.as_mut_ptr() 
  ))?;
  // ...
  libvorbis_return_value_to_result!(vorbis_analysis(
    vorbis_block,
    ptr::null_mut()
  ))?;
  // ...
}
\end{minted}
\vspace{-18pt}
\begin{minted}{c}
int vorbis_block_init(vorbis_dsp_state *v, vorbis_block *vb){
  int i;
  memset(vb,0,sizeof(*vb));
  vb->vd=v;
  // ...
  vorbis_block_internal *vbi=
  vb->internal=_ogg_calloc(1,sizeof(vorbis_block_internal));
  // ...
  vbi->packetblob[i]=&vb->opb;
  // ...
}
\end{minted}
\vspace{-18pt}
\begin{minted}{c}
int vorbis_analysis(vorbis_block *vb, ogg_packet *op){
  // ...
  oggpack_reset(vbi->packetblob[i]);
  // ...
}
\end{minted}
\caption{AXM violation in \texttt{vorbis\_rs}. Top: Rust code
that invokes the library to initialize Rust-owned data structures, followed
by more invocations that pass those data structures as arguments.
Middle: Initialization code which stores in
those data structures pointers to one another. Bottom: Subsequent dereferencing
of those pointers in the library.}
\label{lst:vorbis}
\end{listing}

\section{Related Work}

\nparagraph{Dynamic bug detection in Rust}
ERASan~\cite{minERASANEfficientRust2024} and
RustSan~\cite{choRustSanRetrofittingAddressSanitizer2024}
improve the performance of AddressSanitizer~\cite{serebryanyAddressSanitizerFastAddress2012}
on Rust code by
removing unnecessary run-time checks already
statically guaranteed by Rust principles.
Isolation-based methods for protecting safe Rust code
identify potentially unsafe memory accesses and
isolate their executions through techniques
such as Intel MPK~\cite{bangTRustCompilationFramework2023,kayondoMETASAFECompilingProtecting2024,kirthPKRUsafeAutomaticallyLocking2022,gulmezFriendFoeExploring2023,riveraKeepingSafeRust2021},
or software fault isolation~\cite{lamowskiSandcrustAutomaticSandboxing2017,almohriFideliusCharmIsolating2018,liuSecuringUnsafeRust2020}.
Those methods focus on memory safety violations and
do not identify Rust principle violations in general.
Miri~\cite{RustlangMiri2024} is the official Rust undefined behaviour detection tool
based on interpreting Rust programs at the MIR level.
It detects a wide range of undefined behaviours including violations of Rust principles,
but does not support FFI or inline assembly due to its reliance on high-level
program semantics in MIR.

\nparagraph{Cross-language bug detection}
FFIChecker~\cite{liDetectingCrosslanguageMemory2022} uses
LLVM-IR-level
static taint tracking to find bugs in Rust programs across FFI.
It only covers memory leaks
and mismatching use of heap allocators,
rather than
Rust principle violations.
Its use of LLVM IR also requires recompilation of all external code
with LLVM-based toolchains
and prevents it from handling assembly code or
dynamic linking, unlike \codename{}.
MiriLLI~\cite{mccormackStudyUndefinedBehavior2024} extends the LLVM IR interpreter
to emulate the ABIs of FFI calls
at the LLVM IR level and call back
corresponding Miri functions to perform checks.
Due to reusing Miri,
MiriLLI covers more varieties of undefined
behaviours (e.g., type confusions) than \codename{}.
On the other hand, \codename{}
detects Rust principle violations, a subset of undefined behaviours
unique to Rust, through a simple and general abstraction 
entirely at the machine code level.
MiriLLI shares the same limitations as
FFIChecker due to using LLVM IR.

\nparagraph{Rust aliasing models}
Rust aliasing models formally define valid ways for
references and pointers in Rust to alias with one another
during run-time.
They are formalizations of Rust principles
that also cover mixed Rust code.
Two common aliasing models are Stacked Borrows~\cite{jungStackedBorrowsAliasing2020}
and Tree Borrows~\cite{villaniTreeBorrows2023}.
Although
the latter is generally stricter (i.e., allowing fewer
undefined behaviours) than the former,
both models have significant overlap.
\codename{} is not a proposal of a 
new Rust aliasing model.
It is a design for detecting Rust principle
violations in line with existing aliasing models.

\nparagraph{Capability-based architectures}
Hardware capabilities have a long history~\cite{levyCapabilitybasedComputerSystems1984}.
The more recent capability-based architecture designs include
Mondrian~\cite{witchelMondrianMemoryProtection2002},
CHERI~\cite{watsonCHERIHybridCapabilitySystem2015}, and \capstone{}~\cite{yuCapstoneCapabilitybasedFoundation2023a}.
Existing work has explored using capability-based architectures for
spatial memory safety~\cite{deviettiHardboundArchitecturalSupport2008,witchelMondrianMemoryProtection2002,watsonCHERIHybridCapabilitySystem2015,filloMMachineMulticomputer1995,yuCapstoneCapabilitybasedFoundation2023a},
temporal memory safety~\cite{xiaCHERIvokeCharacterisingPointer2019,filardoCornucopiaReloadedLoad2024,wesleyfilardoCornucopiaTemporalSafety2020,skorstengaardStkTokensEnforcingWellbracketed2019}, and
isolation~\cite{vilanovaCODOMsProtectingSoftware2014,watsonCHERIHybridCapabilitySystem2015,yuCapstoneCapabilitybasedFoundation2023a,esswoodCheriOSDesigningUntrusted2020}.
Most has focused on protecting legacy systems in C/C++.
Recent work proposes using capability-based architectures for unsafe
Rust to safeguard memory
accesses~\cite{harrisRustMorelloAlwaysMemory2023,simStrengtheningMemorySafety2020}.
Those methods only provide protection for memory safety
without covering other types of bugs in Rust.
\capstone{}~\cite{yuCapstoneCapabilitybasedFoundation2023a}
enhances the existing
capability-based models by incorporating extra capability types such as linear capabilities
and revocation capabilities.
Those additions support enclave-like exclusive access guarantees
and revocable capability delegations.

\section{Limitations and Future Work}

Our current \codename{} implementation based on QEMU demonstrates
the practicality of using capabilities for detecting
Rust principle violations at the architecture level through the
revoke-on-use mechanism.
We plan to
investigate how to enable
efficient implementation closer to real hardware in the future.
Revoke-on-use is conceptually compatible with existing use cases of hardware capabilities, but the concrete strategy for integrating those use cases together
requires more investigation.

\section{Conclusions}

We present a novel
revoke-on-use mechanism
for using a hardware
capability-based abstraction
to
enforce Rust principles
across languages at the machine code level.
\codename{}, our design based on revoke-on-use,
yields a readily usable tool
for detecting Rust principle violations
in real-world Rust projects, including those that use
FFI or inline assembly.

\section*{Acknowledgements}
We thank NUS KISP Lab members for their feedback.
This research is supported by a Singapore Ministry of Education (MOE) Tier 2 grant MOE-T2EP20124-0007.

\bibliographystyle{ACM-Reference-Format}
\bibliography{paper,zotero}


\begin{thebibliography}{43}


\ifx \showCODEN    \undefined \def \showCODEN     #1{\unskip}     \fi
\ifx \showISBNx    \undefined \def \showISBNx     #1{\unskip}     \fi
\ifx \showISBNxiii \undefined \def \showISBNxiii  #1{\unskip}     \fi
\ifx \showISSN     \undefined \def \showISSN      #1{\unskip}     \fi
\ifx \showLCCN     \undefined \def \showLCCN      #1{\unskip}     \fi
\ifx \shownote     \undefined \def \shownote      #1{#1}          \fi
\ifx \showarticletitle \undefined \def \showarticletitle #1{#1}   \fi
\ifx \showURL      \undefined \def \showURL       {\relax}        \fi
\providecommand\bibfield[2]{#2}
\providecommand\bibinfo[2]{#2}
\providecommand\natexlab[1]{#1}
\providecommand\showeprint[2][]{arXiv:#2}

\bibitem[Rus({[n.\,d.]})]%
        {RustSecRustSecAdvisory}
 \bibinfo{year}{[n.\,d.]}\natexlab{}.
\newblock \bibinfo{title}{About {{RustSec}} {\guilsinglright} {{RustSec
  Advisory Database}}}.
\newblock \bibinfo{howpublished}{https://rustsec.org/}.
\newblock


\bibitem[Cra({[n.\,d.]})]%
        {CratesioRustPackage}
 \bibinfo{year}{[n.\,d.]}\natexlab{}.
\newblock \bibinfo{title}{Crates.Io: {{Rust Package Registry}}}.
\newblock \bibinfo{howpublished}{https://crates.io/}.
\newblock


\bibitem[Sup(2023)]%
        {SupportingUseRust2023}
 \bibinfo{year}{2023}\natexlab{}.
\newblock \bibinfo{title}{Supporting the {{Use}} of {{Rust}} in the {{Chromium
  Project}}}.
\newblock


\bibitem[Rus(2024)]%
        {RustlangMiri2024}
 \bibinfo{year}{2024}\natexlab{}.
\newblock \bibinfo{title}{Rust-Lang/Miri}.
\newblock \bibinfo{howpublished}{The Rust Programming Language}.
\newblock


\bibitem[Almohri and Evans(2018)]%
        {almohriFideliusCharmIsolating2018}
\bibfield{author}{\bibinfo{person}{Hussain M.~J. Almohri} {and}
  \bibinfo{person}{David Evans}.} \bibinfo{year}{2018}\natexlab{}.
\newblock \showarticletitle{Fidelius {{Charm}}: {{Isolating Unsafe Rust
  Code}}}. In \bibinfo{booktitle}{\emph{Proceedings of the {{Eighth ACM
  Conference}} on {{Data}} and {{Application Security}} and {{Privacy}}}}.
  \bibinfo{publisher}{ACM}, \bibinfo{address}{Tempe AZ USA},
  \bibinfo{pages}{248--255}.
\newblock
\showISBNx{978-1-4503-5632-9}
\href{https://doi.org/10.1145/3176258.3176330}{doi:\nolinkurl{10.1145/3176258.3176330}}


\bibitem[Astrauskas et~al\mbox{.}(2020)]%
        {astrauskasHowProgrammersUse2020}
\bibfield{author}{\bibinfo{person}{Vytautas Astrauskas},
  \bibinfo{person}{Christoph Matheja}, \bibinfo{person}{Federico Poli},
  \bibinfo{person}{Peter M{\"u}ller}, {and} \bibinfo{person}{Alexander~J.
  Summers}.} \bibinfo{year}{2020}\natexlab{}.
\newblock \showarticletitle{How Do Programmers Use Unsafe Rust?}
\newblock \bibinfo{journal}{\emph{Proceedings of the ACM on Programming
  Languages}} \bibinfo{volume}{4}, \bibinfo{number}{OOPSLA}
  (\bibinfo{date}{Nov.} \bibinfo{year}{2020}), \bibinfo{pages}{1--27}.
\newblock
\showISSN{2475-1421}
\href{https://doi.org/10.1145/3428204}{doi:\nolinkurl{10.1145/3428204}}


\bibitem[Bang et~al\mbox{.}(2023)]%
        {bangTRustCompilationFramework2023}
\bibfield{author}{\bibinfo{person}{Inyoung Bang}, \bibinfo{person}{Martin
  Kayondo}, \bibinfo{person}{HyunGon Moon}, {and} \bibinfo{person}{Yunheung
  Paek}.} \bibinfo{year}{2023}\natexlab{}.
\newblock \showarticletitle{{{TRust}}: A Compilation Framework for in-Process
  Isolation to Protect Safe Rust against Untrusted Code}. In
  \bibinfo{booktitle}{\emph{32nd {{USENIX}} Security Symposium ({{USENIX}}
  Security 23)}}. \bibinfo{publisher}{USENIX Association},
  \bibinfo{address}{Anaheim, CA}, \bibinfo{pages}{6947--6964}.
\newblock
\showISBNx{978-1-939133-37-3}


\bibitem[Bellard(2005)]%
        {bellardQEMUFastPortable2005}
\bibfield{author}{\bibinfo{person}{Fabrice Bellard}.}
  \bibinfo{year}{2005}\natexlab{}.
\newblock \showarticletitle{{{QEMU}}, a Fast and Portable Dynamic Translator}.
  In \bibinfo{booktitle}{\emph{2005 {{USENIX}} Annual Technical Conference
  ({{USENIX ATC}} 05)}}. \bibinfo{publisher}{USENIX Association},
  \bibinfo{address}{Anaheim, CA}.
\newblock


\bibitem[Cho et~al\mbox{.}(2024)]%
        {choRustSanRetrofittingAddressSanitizer2024}
\bibfield{author}{\bibinfo{person}{Kyuwon Cho}, \bibinfo{person}{Jongyoon Kim},
  \bibinfo{person}{Kha~Dinh Duy}, \bibinfo{person}{Hajeong Lim}, {and}
  \bibinfo{person}{Hojoon Lee}.} \bibinfo{year}{2024}\natexlab{}.
\newblock \showarticletitle{{{RustSan}}: {{Retrofitting AddressSanitizer}} for
  Efficient Sanitization of Rust}. In \bibinfo{booktitle}{\emph{33rd {{USENIX}}
  Security Symposium ({{USENIX}} Security 24)}}. \bibinfo{publisher}{USENIX
  Association}, \bibinfo{address}{Philadelphia, PA},
  \bibinfo{pages}{3729--3746}.
\newblock
\showISBNx{978-1-939133-44-1}


\bibitem[Claburn(2023)]%
        {claburnMicrosoftRewritingCore2023}
\bibfield{author}{\bibinfo{person}{Thomas Claburn}.}
  \bibinfo{year}{2023}\natexlab{}.
\newblock \bibinfo{title}{Microsoft Is Rewriting Core {{Windows}} Libraries in
  {{Rust}}}.
\newblock
  \bibinfo{howpublished}{https://www.theregister.com/2023/04/27/microsoft\_windows\_rust/}.
\newblock


\bibitem[Devietti et~al\mbox{.}(2008)]%
        {deviettiHardboundArchitecturalSupport2008}
\bibfield{author}{\bibinfo{person}{Joe Devietti}, \bibinfo{person}{Colin
  Blundell}, \bibinfo{person}{Milo M.~K. Martin}, {and} \bibinfo{person}{Steve
  Zdancewic}.} \bibinfo{year}{2008}\natexlab{}.
\newblock \showarticletitle{Hardbound: Architectural Support for Spatial Safety
  of the {{C}} Programming Language}.
\newblock \bibinfo{journal}{\emph{ACM SIGARCH Computer Architecture News}}
  \bibinfo{volume}{36}, \bibinfo{number}{1} (\bibinfo{date}{March}
  \bibinfo{year}{2008}), \bibinfo{pages}{103--114}.
\newblock
\showISSN{0163-5964}
\href{https://doi.org/10.1145/1353534.1346295}{doi:\nolinkurl{10.1145/1353534.1346295}}


\bibitem[Esswood(2020)]%
        {esswoodCheriOSDesigningUntrusted2020}
\bibfield{author}{\bibinfo{person}{Lawrence Esswood}.}
  \bibinfo{year}{2020}\natexlab{}.
\newblock \emph{\bibinfo{title}{{{CheriOS}}: {{Designing}} an Untrusted
  Single-Address-Space Capability Operating System Utilising Capability
  Hardware and a Minimal Hypervisor}}.
\newblock \bibinfo{thesistype}{Ph.\,D. Dissertation}. \bibinfo{school}{Apollo -
  University of Cambridge Repository}.
\newblock
\href{https://doi.org/10.17863/CAM.74163}{doi:\nolinkurl{10.17863/CAM.74163}}


\bibitem[Filardo et~al\mbox{.}(2024)]%
        {filardoCornucopiaReloadedLoad2024}
\bibfield{author}{\bibinfo{person}{Nathaniel~Wesley Filardo},
  \bibinfo{person}{Brett~F. Gutstein}, \bibinfo{person}{Jonathan Woodruff},
  \bibinfo{person}{Jessica Clarke}, \bibinfo{person}{Peter Rugg},
  \bibinfo{person}{Brooks Davis}, \bibinfo{person}{Mark Johnston},
  \bibinfo{person}{Robert Norton}, \bibinfo{person}{David Chisnall},
  \bibinfo{person}{Simon~W. Moore}, \bibinfo{person}{Peter~G. Neumann}, {and}
  \bibinfo{person}{Robert N.~M. Watson}.} \bibinfo{year}{2024}\natexlab{}.
\newblock \showarticletitle{Cornucopia {{Reloaded}}: {{Load Barriers}} for
  {{CHERI Heap Temporal Safety}}}. In \bibinfo{booktitle}{\emph{Proceedings of
  the 29th {{ACM International Conference}} on {{Architectural Support}} for
  {{Programming Languages}} and {{Operating Systems}}, {{Volume}} 2}}.
  \bibinfo{publisher}{ACM}, \bibinfo{address}{La Jolla CA USA},
  \bibinfo{pages}{251--268}.
\newblock
\showISBNx{979-8-4007-0385-0}
\href{https://doi.org/10.1145/3620665.3640416}{doi:\nolinkurl{10.1145/3620665.3640416}}


\bibitem[Fillo et~al\mbox{.}(1995)]%
        {filloMMachineMulticomputer1995}
\bibfield{author}{\bibinfo{person}{M. Fillo}, \bibinfo{person}{S.W. Keckler},
  \bibinfo{person}{W.J. Dally}, \bibinfo{person}{N.P. Carter},
  \bibinfo{person}{A. Chang}, \bibinfo{person}{Y. Gurevich}, {and}
  \bibinfo{person}{W.S. Lee}.} \bibinfo{year}{1995}\natexlab{}.
\newblock \showarticletitle{The {{M-Machine}} Multicomputer}. In
  \bibinfo{booktitle}{\emph{Proceedings of the 28th {{Annual International
  Symposium}} on {{Microarchitecture}}}}. \bibinfo{publisher}{IEEE},
  \bibinfo{address}{Ann Arbor, MI, USA}, \bibinfo{pages}{146--156}.
\newblock
\showISBNx{978-0-8186-7349-8}
\href{https://doi.org/10.1109/MICRO.1995.476822}{doi:\nolinkurl{10.1109/MICRO.1995.476822}}


\bibitem[G{\"u}lmez et~al\mbox{.}(2023)]%
        {gulmezFriendFoeExploring2023}
\bibfield{author}{\bibinfo{person}{Merve G{\"u}lmez}, \bibinfo{person}{Thomas
  Nyman}, \bibinfo{person}{Christoph Baumann}, {and}
  \bibinfo{person}{Jan~Tobias M{\"u}hlberg}.} \bibinfo{year}{2023}\natexlab{}.
\newblock \bibinfo{title}{Friend or {{Foe Inside}}? {{Exploring In-Process
  Isolation}} to {{Maintain Memory Safety}} for {{Unsafe Rust}}}.
\newblock
\showeprint[arxiv]{2306.08127}~[cs]


\bibitem[Harris et~al\mbox{.}(2023)]%
        {harrisRustMorelloAlwaysMemory2023}
\bibfield{author}{\bibinfo{person}{Sarah Harris}, \bibinfo{person}{Simon
  Cooksey}, \bibinfo{person}{Michael Vollmer}, {and} \bibinfo{person}{Mark
  Batty}.} \bibinfo{year}{2023}\natexlab{}.
\newblock \showarticletitle{Rust for {{Morello}}: {{Always-On Memory Safety}},
  {{Even}} in {{Unsafe Code}}}. In \bibinfo{booktitle}{\emph{37th {{European
  Conference}} on {{Object-Oriented Programming}} ({{ECOOP}} 2023)}}
  \emph{(\bibinfo{series}{Leibniz {{International Proceedings}} in
  {{Informatics}} ({{LIPIcs}})}, Vol.~\bibinfo{volume}{263})},
  \bibfield{editor}{\bibinfo{person}{Karim Ali} {and} \bibinfo{person}{Guido
  Salvaneschi}} (Eds.). \bibinfo{publisher}{Schloss Dagstuhl -- Leibniz-Zentrum
  f{\"u}r Informatik}, \bibinfo{address}{Dagstuhl, Germany},
  \bibinfo{pages}{39:1--39:27}.
\newblock
\showISBNx{978-3-95977-281-5}
\showISSN{1868-8969}
\href{https://doi.org/10.4230/LIPIcs.ECOOP.2023.39}{doi:\nolinkurl{10.4230/LIPIcs.ECOOP.2023.39}}


\bibitem[{Jonathan Corbet}(2022)]%
        {jonathancorbetFirstLookRust2022}
\bibfield{author}{\bibinfo{person}{{Jonathan Corbet}}.}
  \bibinfo{year}{2022}\natexlab{}.
\newblock \bibinfo{title}{A First Look at {{Rust}} in the 6.1 Kernel
  [{{LWN}}.Net]}.
\newblock \bibinfo{howpublished}{https://lwn.net/Articles/910762/}.
\newblock


\bibitem[Jung et~al\mbox{.}(2020)]%
        {jungStackedBorrowsAliasing2020}
\bibfield{author}{\bibinfo{person}{Ralf Jung}, \bibinfo{person}{Hoang-Hai
  Dang}, \bibinfo{person}{Jeehoon Kang}, {and} \bibinfo{person}{Derek Dreyer}.}
  \bibinfo{year}{2020}\natexlab{}.
\newblock \showarticletitle{Stacked Borrows: An Aliasing Model for {{Rust}}}.
\newblock \bibinfo{journal}{\emph{Proceedings of the ACM on Programming
  Languages}} \bibinfo{volume}{4}, \bibinfo{number}{POPL} (\bibinfo{date}{Jan.}
  \bibinfo{year}{2020}), \bibinfo{pages}{1--32}.
\newblock
\showISSN{2475-1421}
\href{https://doi.org/10.1145/3371109}{doi:\nolinkurl{10.1145/3371109}}


\bibitem[Kayondo et~al\mbox{.}(2024)]%
        {kayondoMETASAFECompilingProtecting2024}
\bibfield{author}{\bibinfo{person}{Martin Kayondo}, \bibinfo{person}{Inyoung
  Bang}, \bibinfo{person}{Yeongjun Kwak}, \bibinfo{person}{Hyungon Moon}, {and}
  \bibinfo{person}{Yunheung Paek}.} \bibinfo{year}{2024}\natexlab{}.
\newblock \showarticletitle{{{METASAFE}}: {{Compiling}} for {{Protecting Smart
  Pointer Metadata}} to {{Ensure Safe Rust Integrity}}}. In
  \bibinfo{booktitle}{\emph{{{USENIX Security}} '24}}.
\newblock


\bibitem[Kirth et~al\mbox{.}(2022)]%
        {kirthPKRUsafeAutomaticallyLocking2022}
\bibfield{author}{\bibinfo{person}{Paul Kirth}, \bibinfo{person}{Mitchel
  Dickerson}, \bibinfo{person}{Stephen Crane}, \bibinfo{person}{Per Larsen},
  \bibinfo{person}{Adrian Dabrowski}, \bibinfo{person}{David Gens},
  \bibinfo{person}{Yeoul Na}, \bibinfo{person}{Stijn Volckaert}, {and}
  \bibinfo{person}{Michael Franz}.} \bibinfo{year}{2022}\natexlab{}.
\newblock \showarticletitle{{{PKRU-safe}}: Automatically Locking down the Heap
  between Safe and Unsafe Languages}. In \bibinfo{booktitle}{\emph{Proceedings
  of the {{Seventeenth European Conference}} on {{Computer Systems}}}}.
  \bibinfo{publisher}{ACM}, \bibinfo{address}{Rennes France},
  \bibinfo{pages}{132--148}.
\newblock
\showISBNx{978-1-4503-9162-7}
\href{https://doi.org/10.1145/3492321.3519582}{doi:\nolinkurl{10.1145/3492321.3519582}}


\bibitem[Lamowski et~al\mbox{.}(2017)]%
        {lamowskiSandcrustAutomaticSandboxing2017}
\bibfield{author}{\bibinfo{person}{Benjamin Lamowski}, \bibinfo{person}{Carsten
  Weinhold}, \bibinfo{person}{Adam Lackorzynski}, {and}
  \bibinfo{person}{Hermann H{\"a}rtig}.} \bibinfo{year}{2017}\natexlab{}.
\newblock \showarticletitle{Sandcrust: {{Automatic Sandboxing}} of {{Unsafe
  Components}} in {{Rust}}}. In \bibinfo{booktitle}{\emph{Proceedings of the
  9th {{Workshop}} on {{Programming Languages}} and {{Operating Systems}}}}.
  \bibinfo{publisher}{ACM}, \bibinfo{address}{Shanghai China},
  \bibinfo{pages}{51--57}.
\newblock
\showISBNx{978-1-4503-5153-9}
\href{https://doi.org/10.1145/3144555.3144562}{doi:\nolinkurl{10.1145/3144555.3144562}}


\bibitem[Levy(1984)]%
        {levyCapabilitybasedComputerSystems1984}
\bibfield{author}{\bibinfo{person}{Henry~M. Levy}.}
  \bibinfo{year}{1984}\natexlab{}.
\newblock \bibinfo{booktitle}{\emph{Capability-Based Computer Systems}}.
\newblock \bibinfo{publisher}{Digital Press}, \bibinfo{address}{Bedford, Mass}.
\newblock
\showISBNx{978-0-932376-22-0}


\bibitem[Li et~al\mbox{.}(2022)]%
        {liDetectingCrosslanguageMemory2022}
\bibfield{author}{\bibinfo{person}{Zhuohua Li}, \bibinfo{person}{Jincheng
  Wang}, \bibinfo{person}{Mingshen Sun}, {and} \bibinfo{person}{John C.~S.
  Lui}.} \bibinfo{year}{2022}\natexlab{}.
\newblock \showarticletitle{Detecting {{Cross-language Memory Management
  Issues}} in {{Rust}}}.
\newblock In \bibinfo{booktitle}{\emph{Computer {{Security}} -- {{ESORICS}}
  2022}}, \bibfield{editor}{\bibinfo{person}{Vijayalakshmi Atluri},
  \bibinfo{person}{Roberto Di~Pietro}, \bibinfo{person}{Christian~D. Jensen},
  {and} \bibinfo{person}{Weizhi Meng}} (Eds.). Vol.~\bibinfo{volume}{13556}.
  \bibinfo{publisher}{Springer Nature Switzerland}, \bibinfo{address}{Cham},
  \bibinfo{pages}{680--700}.
\newblock
\showISBNx{978-3-031-17142-0 978-3-031-17143-7}
\href{https://doi.org/10.1007/978-3-031-17143-7_33}{doi:\nolinkurl{10.1007/978-3-031-17143-7_33}}


\bibitem[Liu et~al\mbox{.}(2020)]%
        {liuSecuringUnsafeRust2020}
\bibfield{author}{\bibinfo{person}{Peiming Liu}, \bibinfo{person}{Gang Zhao},
  {and} \bibinfo{person}{Jeff Huang}.} \bibinfo{year}{2020}\natexlab{}.
\newblock \showarticletitle{Securing Unsafe Rust Programs with {{XRust}}}. In
  \bibinfo{booktitle}{\emph{Proceedings of the {{ACM}}/{{IEEE}} 42nd
  {{International Conference}} on {{Software Engineering}}}}.
  \bibinfo{publisher}{ACM}, \bibinfo{address}{Seoul South Korea},
  \bibinfo{pages}{234--245}.
\newblock
\showISBNx{978-1-4503-7121-6}
\href{https://doi.org/10.1145/3377811.3380325}{doi:\nolinkurl{10.1145/3377811.3380325}}


\bibitem[McCormack et~al\mbox{.}(2024)]%
        {mccormackStudyUndefinedBehavior2024}
\bibfield{author}{\bibinfo{person}{Ian McCormack}, \bibinfo{person}{Joshua
  Sunshine}, {and} \bibinfo{person}{Jonathan Aldrich}.}
  \bibinfo{year}{2024}\natexlab{}.
\newblock \bibinfo{title}{A {{Study}} of {{Undefined Behavior Across Foreign
  Function Boundaries}} in {{Rust Libraries}}}.
\newblock
\showeprint[arxiv]{2404.11671}~[cs]


\bibitem[Min et~al\mbox{.}(2024)]%
        {minERASANEfficientRust2024}
\bibfield{author}{\bibinfo{person}{J. Min}, \bibinfo{person}{D. Yu},
  \bibinfo{person}{S. Jeong}, \bibinfo{person}{D. Song}, {and}
  \bibinfo{person}{Y. Jeon}.} \bibinfo{year}{2024}\natexlab{}.
\newblock \showarticletitle{{{ERASAN}} : {{Efficient}} Rust Address Sanitizer}.
  In \bibinfo{booktitle}{\emph{2024 {{IEEE}} Symposium on Security and Privacy
  ({{SP}})}}. \bibinfo{publisher}{IEEE Computer Society}, \bibinfo{address}{Los
  Alamitos, CA, USA}, \bibinfo{pages}{239--239}.
\newblock
\showISSN{2375-1207}
\href{https://doi.org/10.1109/SP54263.2024.00182}{doi:\nolinkurl{10.1109/SP54263.2024.00182}}


\bibitem[Nagarakatte et~al\mbox{.}(2009)]%
        {nagarakatteSoftBoundHighlyCompatible2009}
\bibfield{author}{\bibinfo{person}{Santosh Nagarakatte},
  \bibinfo{person}{Jianzhou Zhao}, \bibinfo{person}{Milo~M.K. Martin}, {and}
  \bibinfo{person}{Steve Zdancewic}.} \bibinfo{year}{2009}\natexlab{}.
\newblock \showarticletitle{{{SoftBound}}: Highly Compatible and Complete
  Spatial Memory Safety for c}.
\newblock \bibinfo{journal}{\emph{ACM SIGPLAN Notices}} \bibinfo{volume}{44},
  \bibinfo{number}{6} (\bibinfo{date}{May} \bibinfo{year}{2009}),
  \bibinfo{pages}{245--258}.
\newblock
\showISSN{0362-1340, 1558-1160}
\href{https://doi.org/10.1145/1543135.1542504}{doi:\nolinkurl{10.1145/1543135.1542504}}


\bibitem[Nagarakatte et~al\mbox{.}(2010)]%
        {cets}
\bibfield{author}{\bibinfo{person}{Santosh Nagarakatte},
  \bibinfo{person}{Jianzhou Zhao}, \bibinfo{person}{Milo~M.K. Martin}, {and}
  \bibinfo{person}{Steve Zdancewic}.} \bibinfo{year}{2010}\natexlab{}.
\newblock \showarticletitle{CETS: compiler enforced temporal safety for C}. In
  \bibinfo{booktitle}{\emph{Proceedings of the 2010 International Symposium on
  Memory Management}} (Toronto, Ontario, Canada) \emph{(\bibinfo{series}{ISMM
  '10})}. \bibinfo{publisher}{Association for Computing Machinery},
  \bibinfo{address}{New York, NY, USA}, \bibinfo{pages}{31–40}.
\newblock
\showISBNx{9781450300544}
\href{https://doi.org/10.1145/1806651.1806657}{doi:\nolinkurl{10.1145/1806651.1806657}}


\bibitem[Rivera et~al\mbox{.}(2021)]%
        {riveraKeepingSafeRust2021}
\bibfield{author}{\bibinfo{person}{Elijah Rivera}, \bibinfo{person}{Samuel
  Mergendahl}, \bibinfo{person}{Howard Shrobe}, \bibinfo{person}{Hamed
  Okhravi}, {and} \bibinfo{person}{Nathan Burow}.}
  \bibinfo{year}{2021}\natexlab{}.
\newblock \showarticletitle{Keeping {{Safe Rust Safe}} with {{Galeed}}}. In
  \bibinfo{booktitle}{\emph{Annual {{Computer Security Applications
  Conference}}}}. \bibinfo{publisher}{ACM}, \bibinfo{address}{Virtual Event
  USA}, \bibinfo{pages}{824--836}.
\newblock
\showISBNx{978-1-4503-8579-4}
\href{https://doi.org/10.1145/3485832.3485903}{doi:\nolinkurl{10.1145/3485832.3485903}}


\bibitem[Serebryany et~al\mbox{.}(2012)]%
        {serebryanyAddressSanitizerFastAddress2012}
\bibfield{author}{\bibinfo{person}{Konstantin Serebryany},
  \bibinfo{person}{Derek Bruening}, \bibinfo{person}{Alexander Potapenko},
  {and} \bibinfo{person}{Dmitry Vyukov}.} \bibinfo{year}{2012}\natexlab{}.
\newblock \showarticletitle{{{AddressSanitizer}}: A Fast Address Sanity
  Checker}. In \bibinfo{booktitle}{\emph{Proceedings of the 2012 {{USENIX}}
  Conference on Annual Technical Conference}} \emph{(\bibinfo{series}{Usenix
  Atc'12})}. \bibinfo{publisher}{USENIX Association}, \bibinfo{address}{USA},
  \bibinfo{pages}{28}.
\newblock


\bibitem[Serebryany and Iskhodzhanov(2009)]%
        {serebryanyThreadSanitizerDataRace2009}
\bibfield{author}{\bibinfo{person}{Konstantin Serebryany} {and}
  \bibinfo{person}{Timur Iskhodzhanov}.} \bibinfo{year}{2009}\natexlab{}.
\newblock \showarticletitle{{{ThreadSanitizer}}: Data Race Detection in
  Practice}. In \bibinfo{booktitle}{\emph{Proceedings of the {{Workshop}} on
  {{Binary Instrumentation}} and {{Applications}}}}. \bibinfo{publisher}{ACM},
  \bibinfo{address}{New York New York USA}, \bibinfo{pages}{62--71}.
\newblock
\showISBNx{978-1-60558-793-6}
\href{https://doi.org/10.1145/1791194.1791203}{doi:\nolinkurl{10.1145/1791194.1791203}}


\bibitem[Sim(2020)]%
        {simStrengtheningMemorySafety2020}
\bibfield{author}{\bibinfo{person}{Nicholas Wei~Sheng Sim}.}
  \bibinfo{year}{2020}\natexlab{}.
\newblock \emph{\bibinfo{title}{Strengthening Memory Safety in {{Rust}}:
  Exploring {{CHERI}} Capabilities for a Safe Language}}.
\newblock \bibinfo{thesistype}{Ph.\,D. Dissertation}.
  \bibinfo{school}{University of Cambridge, Computer Laboratory}.
\newblock


\bibitem[Skorstengaard et~al\mbox{.}(2019)]%
        {skorstengaardStkTokensEnforcingWellbracketed2019}
\bibfield{author}{\bibinfo{person}{Lau Skorstengaard},
  \bibinfo{person}{Dominique Devriese}, {and} \bibinfo{person}{Lars Birkedal}.}
  \bibinfo{year}{2019}\natexlab{}.
\newblock \showarticletitle{{{StkTokens}}: Enforcing Well-Bracketed Control
  Flow and Stack Encapsulation Using Linear Capabilities}.
\newblock \bibinfo{journal}{\emph{Proceedings of the ACM on Programming
  Languages}} \bibinfo{volume}{3}, \bibinfo{number}{POPL} (\bibinfo{date}{Jan.}
  \bibinfo{year}{2019}), \bibinfo{pages}{1--28}.
\newblock
\showISSN{2475-1421}
\href{https://doi.org/10.1145/3290332}{doi:\nolinkurl{10.1145/3290332}}


\bibitem[Vilanova et~al\mbox{.}(2014)]%
        {vilanovaCODOMsProtectingSoftware2014}
\bibfield{author}{\bibinfo{person}{Llu{\"i}s Vilanova}, \bibinfo{person}{Muli
  {Ben-Yehuda}}, \bibinfo{person}{Nacho Navarro}, \bibinfo{person}{Yoav
  Etsion}, {and} \bibinfo{person}{Mateo Valero}.}
  \bibinfo{year}{2014}\natexlab{}.
\newblock \showarticletitle{{{CODOMs}}: Protecting Software with Code-Centric
  Memory Domains}.
\newblock \bibinfo{journal}{\emph{ACM SIGARCH Computer Architecture News}}
  \bibinfo{volume}{42}, \bibinfo{number}{3} (\bibinfo{date}{Oct.}
  \bibinfo{year}{2014}), \bibinfo{pages}{469--480}.
\newblock
\showISSN{0163-5964}
\href{https://doi.org/10.1145/2678373.2665741}{doi:\nolinkurl{10.1145/2678373.2665741}}


\bibitem[Villani(2023)]%
        {villaniTreeBorrows2023}
\bibfield{author}{\bibinfo{person}{Neven Villani}.}
  \bibinfo{year}{2023}\natexlab{}.
\newblock \bibinfo{title}{Tree {{Borrows}}}.
\newblock


\bibitem[Waterman et~al\mbox{.}({[n.\,d.]})]%
        {watermanRISCVInstructionSeta}
\bibfield{author}{\bibinfo{person}{Andrew Waterman}, \bibinfo{person}{Krste
  Asanovic}, \bibinfo{person}{John Hauser}, {and} \bibinfo{person}{{\relax CS}
  Division}.} \bibinfo{year}{[n.\,d.]}\natexlab{}.
\newblock \bibinfo{title}{The {{RISC-V Instruction Set Manual}} ({{Volume II}}:
  {{Privileged Architecture}})}.
\newblock


\bibitem[Watson et~al\mbox{.}(2015)]%
        {watsonCHERIHybridCapabilitySystem2015}
\bibfield{author}{\bibinfo{person}{Robert~N.M. Watson},
  \bibinfo{person}{Jonathan Woodruff}, \bibinfo{person}{Peter~G. Neumann},
  \bibinfo{person}{Simon~W. Moore}, \bibinfo{person}{Jonathan Anderson},
  \bibinfo{person}{David Chisnall}, \bibinfo{person}{Nirav Dave},
  \bibinfo{person}{Brooks Davis}, \bibinfo{person}{Khilan Gudka},
  \bibinfo{person}{Ben Laurie}, \bibinfo{person}{Steven~J. Murdoch},
  \bibinfo{person}{Robert Norton}, \bibinfo{person}{Michael Roe},
  \bibinfo{person}{Stacey Son}, {and} \bibinfo{person}{Munraj Vadera}.}
  \bibinfo{year}{2015}\natexlab{}.
\newblock \showarticletitle{{{CHERI}}: {{A Hybrid Capability-System
  Architecture}} for {{Scalable Software Compartmentalization}}}. In
  \bibinfo{booktitle}{\emph{2015 {{IEEE Symposium}} on {{Security}} and
  {{Privacy}}}}. \bibinfo{publisher}{IEEE}, \bibinfo{address}{San Jose, CA},
  \bibinfo{pages}{20--37}.
\newblock
\showISBNx{978-1-4673-6949-7}
\href{https://doi.org/10.1109/SP.2015.9}{doi:\nolinkurl{10.1109/SP.2015.9}}


\bibitem[Wesley~Filardo et~al\mbox{.}(2020)]%
        {wesleyfilardoCornucopiaTemporalSafety2020}
\bibfield{author}{\bibinfo{person}{Nathaniel Wesley~Filardo},
  \bibinfo{person}{Brett~F. Gutstein}, \bibinfo{person}{Jonathan Woodruff},
  \bibinfo{person}{Sam Ainsworth}, \bibinfo{person}{Lucian {Paul-Trifu}},
  \bibinfo{person}{Brooks Davis}, \bibinfo{person}{Hongyan Xia},
  \bibinfo{person}{Edward Tomasz~Napierala}, \bibinfo{person}{Alexander
  Richardson}, \bibinfo{person}{John Baldwin}, \bibinfo{person}{David
  Chisnall}, \bibinfo{person}{Jessica Clarke}, \bibinfo{person}{Khilan Gudka},
  \bibinfo{person}{Alexandre Joannou}, \bibinfo{person}{A. Theodore~Markettos},
  \bibinfo{person}{Alfredo Mazzinghi}, \bibinfo{person}{Robert~M. Norton},
  \bibinfo{person}{Michael Roe}, \bibinfo{person}{Peter Sewell},
  \bibinfo{person}{Stacey Son}, \bibinfo{person}{Timothy~M. Jones},
  \bibinfo{person}{Simon~W. Moore}, \bibinfo{person}{Peter~G. Neumann}, {and}
  \bibinfo{person}{Robert N.~M. Watson}.} \bibinfo{year}{2020}\natexlab{}.
\newblock \showarticletitle{Cornucopia: {{Temporal Safety}} for {{CHERI
  Heaps}}}. In \bibinfo{booktitle}{\emph{2020 {{IEEE Symposium}} on
  {{Security}} and {{Privacy}} ({{SP}})}}. \bibinfo{publisher}{IEEE},
  \bibinfo{address}{San Francisco, CA, USA}, \bibinfo{pages}{608--625}.
\newblock
\showISBNx{978-1-7281-3497-0}
\href{https://doi.org/10.1109/SP40000.2020.00098}{doi:\nolinkurl{10.1109/SP40000.2020.00098}}


\bibitem[Witchel et~al\mbox{.}(2002)]%
        {witchelMondrianMemoryProtection2002}
\bibfield{author}{\bibinfo{person}{Emmett Witchel}, \bibinfo{person}{Josh
  Cates}, {and} \bibinfo{person}{Krste Asanovi{\'c}}.}
  \bibinfo{year}{2002}\natexlab{}.
\newblock \showarticletitle{Mondrian Memory Protection}. In
  \bibinfo{booktitle}{\emph{Proceedings of the 10th International Conference on
  {{Architectural}} Support for Programming Languages and Operating Systems}}.
  \bibinfo{publisher}{ACM}, \bibinfo{address}{San Jose California},
  \bibinfo{pages}{304--316}.
\newblock
\showISBNx{978-1-58113-574-9}
\href{https://doi.org/10.1145/605397.605429}{doi:\nolinkurl{10.1145/605397.605429}}


\bibitem[Xia et~al\mbox{.}(2019)]%
        {xiaCHERIvokeCharacterisingPointer2019}
\bibfield{author}{\bibinfo{person}{Hongyan Xia}, \bibinfo{person}{Jonathan
  Woodruff}, \bibinfo{person}{Sam Ainsworth}, \bibinfo{person}{Nathaniel~W.
  Filardo}, \bibinfo{person}{Michael Roe}, \bibinfo{person}{Alexander
  Richardson}, \bibinfo{person}{Peter Rugg}, \bibinfo{person}{Peter~G.
  Neumann}, \bibinfo{person}{Simon~W. Moore}, \bibinfo{person}{Robert N.~M.
  Watson}, {and} \bibinfo{person}{Timothy~M. Jones}.}
  \bibinfo{year}{2019}\natexlab{}.
\newblock \showarticletitle{{{CHERIvoke}}: {{Characterising Pointer
  Revocation}} Using {{CHERI Capabilities}} for {{Temporal Memory Safety}}}. In
  \bibinfo{booktitle}{\emph{Proceedings of the 52nd {{Annual IEEE}}/{{ACM
  International Symposium}} on {{Microarchitecture}}}}.
  \bibinfo{publisher}{ACM}, \bibinfo{address}{Columbus OH USA},
  \bibinfo{pages}{545--557}.
\newblock
\showISBNx{978-1-4503-6938-1}
\href{https://doi.org/10.1145/3352460.3358288}{doi:\nolinkurl{10.1145/3352460.3358288}}


\bibitem[Yu et~al\mbox{.}(2023)]%
        {yuCapstoneCapabilitybasedFoundation2023a}
\bibfield{author}{\bibinfo{person}{Jason~Zhijingcheng Yu},
  \bibinfo{person}{Conrad Watt}, \bibinfo{person}{Aditya Badole},
  \bibinfo{person}{Trevor~E. Carlson}, {and} \bibinfo{person}{Prateek Saxena}.}
  \bibinfo{year}{2023}\natexlab{}.
\newblock \showarticletitle{Capstone: {{A Capability-based Foundation}} for
  {{Trustless Secure Memory Access}}}. In \bibinfo{booktitle}{\emph{32nd
  {{USENIX Security Symposium}} ({{USENIX Security}} 23)}}.
  \bibinfo{publisher}{USENIX Association}, \bibinfo{address}{Anaheim, CA},
  \bibinfo{pages}{787--804}.
\newblock
\showISBNx{978-1-939133-37-3}


\bibitem[Yu et~al\mbox{.}(2025)]%
        {yuArtifactsSecuringMixed2025}
\bibfield{author}{\bibinfo{person}{Zhijingcheng Yu}, \bibinfo{person}{Fangqi
  Han}, \bibinfo{person}{Kaustab Choudhury}, \bibinfo{person}{Trevor~E.
  Carlson}, {and} \bibinfo{person}{Prateek Saxena}.}
  \bibinfo{year}{2025}\natexlab{}.
\newblock \bibinfo{title}{Artifacts of "{{Securing Mixed Rust}} with {{Hardware
  Capabilities}}"}.
\newblock \bibinfo{howpublished}{Zenodo}.
\newblock
\href{https://doi.org/10.5281/ZENODO.14625327}{doi:\nolinkurl{10.5281/ZENODO.14625327}}


\bibitem[Zhang et~al\mbox{.}(2023)]%
        {zhangDualNatureNecessity2023}
\bibfield{author}{\bibinfo{person}{Yuchen Zhang}, \bibinfo{person}{Ashish
  Kundu}, \bibinfo{person}{Georgios Portokalidis}, {and} \bibinfo{person}{Jun
  Xu}.} \bibinfo{year}{2023}\natexlab{}.
\newblock \showarticletitle{On the {{Dual Nature}} of {{Necessity}} in {{Use}}
  of {{Rust Unsafe Code}}}. In \bibinfo{booktitle}{\emph{Proceedings of the
  31st {{ACM Joint European Software Engineering Conference}} and {{Symposium}}
  on the {{Foundations}} of {{Software Engineering}}}}.
  \bibinfo{publisher}{ACM}, \bibinfo{address}{San Francisco CA USA},
  \bibinfo{pages}{2032--2037}.
\newblock
\showISBNx{979-8-4007-0327-0}
\href{https://doi.org/10.1145/3611643.3613878}{doi:\nolinkurl{10.1145/3611643.3613878}}


\end{thebibliography}

\appendix

\section{Additional Details in Design}
\label{appx:aux-def}

The \codename{} instruction semantics 
in Section~\ref{subsec:instructions} references
the following auxiliary definitions.

Let $c = m(i)$ be a capability. We use $C_m(i)$
to denote the set of identifiers of all children of $c$ in
its borrow tree:
\begin{equation}
C_m(i) = \{i^\prime \mid
m(i^\prime) = (a^\prime_l, a^\prime_l, p^\prime, i)\}.
\end{equation}
Then we define the capability derived set in $m$, which contains identifiers of all capabilities
in a subtree, written $\derived_m(i)$:
\begin{equation}
\derived_m(i) = \{i\} \cup \bigcup_{i^\prime \in C_m(i)} \derived_m(i^\prime).
\end{equation}
When $S$ is a set of capability identifiers,
$\derived_m(S) = \bigcup_{i \in S}\derived_m(i)$.

$\overlap_m(a)$ is the set of identifiers of capabilities that overlap with
$a$:
\begin{equation}
\overlap_m(a) = \{ i \mid m(i) = (a_l, a_r, p, b), \quad a_l \leq a < a_r \}.
\end{equation}

$\revoke(m, S)$ is a new capability map that is the same as $m$ except that all
capabilities with identifiers in $S$ are invalid:
\begin{equation}
\revoke(m, S) = \{i \mapsto \revokeperm(c, i, S) \mid
m(i) = c\},
\end{equation}
where
\begin{equation}
\revokeperm((a_l, a_r, p, b), i, S) = \begin{cases}
(a_l, a_r, p, b) & \text{if } i \in S \\
(a_l, a_r, \text{NA}, b) & \text{otherwise}.
\end{cases}
\end{equation}

$\revoker(m, S)$ demotes capabilities not in $S$ to $\text{RO}$:
\begin{equation}
\revoker(m, S) = \{i \mapsto \revokepermr(c, i, S) \mid
m(i) = c\},
\end{equation}
where
\begin{equation}
\revokepermr((a_l, a_r, p, b), i, S) = \begin{cases}
(a_l, a_r, p, b) & \text{if } i \in S \\
(a_l, a_r, \text{RO}, b) & \text{if } i \not\in S, p = \text{RO} \\
(a_l, a_r, \text{NA}, b) & \text{otherwise}.
\end{cases}
\end{equation}

$R_m(i)$ is the identifier of the root capability of the borrow tree $m(i)$ belongs to:
\begin{equation}
R_m(i) = \begin{cases}
    i & \text{if } m(i) = (a_l, a_r, p, \bot) \\
    R_m(i^\prime) & \text{if } m(i) = (a_l, a_r, p, i^\prime).
\end{cases}
\end{equation}
$A_m(i)$ contains all ancestors of $i$ in the borrow tree:
\begin{equation}
A_m(i) = \begin{cases}
    \{i\} & \text{if } m(i) = (a_l, a_r, p, \bot) \\
    \{i\} \cup A_m(i^\prime) & \text{if } m(i) = (a_l, a_r, p, i^\prime).
\end{cases}
\end{equation}

%




\section{Artifact Appendix}




\subsection{Abstract}

We include as artifacts all implementation code as well as
benchmarks and scripts used in this work.
The implementation half of the artifacts consists mainly of two
components: a modified toolchain for instrumenting mixed Rust code,
and a QEMU-based implementation of the hardware architecture
that runs the instrumented code and detects errors.
The benchmarks and scripts are used in evaluation (Section~\ref{sec:eval}).
Scripts for setting up and running baselines are also included.
The artifacts are licensed under Creative Commons Attribution 4.0 International.



\subsection{Description \& Requirements}



At the top level the artifacts consist of three directories: \texttt{setup},
\texttt{tests}, and \texttt{evaluations}.
\begin{itemize}
    \item \texttt{setup}: scripts and patches for setting up these artifacts.
    \item \texttt{tests}: small example programs used to test the setup and as
        simple demos.
    \item \texttt{evaluations}: data and scripts used in the experiments of this
        work. This directory is in turn organized into subdirectories each
        corresponding to a different part of the evaluation section (Section~\ref{sec:eval}).
\end{itemize}

To help users navigate, we include in most directories
\texttt{README} files that
explain their contents.


\subsubsection{Security, privacy, and ethical concerns}

These artifacts do not expose the user to any risks that concern
security or privacy.
No special privilege or personal information is required to set up or use them.



\subsubsection{How to access}

The artifacts are available on Zenodo with DOI 10.5281/ZENODO.14625327.




\subsubsection{Hardware dependencies}

No significant hardware dependencies. An x86-64 platform is expected.



\subsubsection{Software dependencies}

The artifacts require a Linux environment. Docker is optional
but strongly recommended.



\subsubsection{Benchmarks}

The benchmarks are all collected from the Rust crate registry
\textit{crates.io}.
Both the original benchmarks used in this work and the scripts used
for collecting them are included in the artifacts.
No additional benchmarks need to be present to use these artifacts.


\subsection{Set Up}
\label{subsec:setup}



\subsubsection{Installation}

The artifacts can be set up through either Docker (recommended) or
scripts that run in the host environment. We assume the former option
here. 
To pull the Docker image, run
\begin{Verbatim}
./pull-docker 
\end{Verbatim}
Alternatively,
one can build the Docker image from scratch with the provided
\verb|Dockerfile| by running
\begin{Verbatim}
./build-docker
\end{Verbatim}
For full instructions, please consult the \texttt{README} document.
Roughly 10 GB of disk space is required for the setup.


\subsubsection{Basic test}


The \texttt{tests} directory contains two simple programs to test the setup.
Use
\begin{Verbatim}
 ./docker-test   
\end{Verbatim}
to build and run them in succession.
The first program, \texttt{helloworld}, should run successfully and
print \texttt{Hello, world!}.
The second program, \texttt{violation}, should trigger an error
saying \texttt{Attempting to use an invalid capability for load}.




\subsection{Evaluation Workflow}



\subsubsection{Major claims}


\begin{compactitem}
\item[(C1):] \codename{} is capable of passing over $99\%$ of the test cases
bundled with the top 100 popular Rust crates. This is proven by the experiment
presented in Section~\ref{subsec:compat} with results reported in
Table~\ref{tab:compat-stats} and Figure~\ref{fig:venn-compat}.

\item[(C2):] \codename{} is able to detect existing bugs
caused by Rust principle violations, matching Miri results,
while AddressSanitizer and ThreadSanitizer are unable to detect $40\%$
of them. This is proven by the experiment presented in Section~\ref{subsec:bug-detect}
with results shown in Table~\ref{tab:benchmarks-cves}.
  %
\item[(C3):] \codename{} detects 8 previously-unknown bugs in real-world mixed Rust code, which Miri does not support.
    This is proven by the experiment presented in
    Section~\ref{subsec:bug-new} with results summarized
    in Table~\ref{tab:bugs-new}.
\item[(C4):] \codename{} achieves an average speed-up of over 2x in comparison with Miri. This is proven in
Section~\ref{subsec:perf} with results illustrated in Figure~\ref{fig:performance-hist}.
\end{compactitem}


\subsubsection{Experiments}


We provide scripts for running the experiments at the top level
of the artifact directory. Assuming that Docker is used, to run each experiment,
simply
use
\begin{Verbatim}[breaklines=true]
./docker-eval<n> [nprocs]
\end{Verbatim}
where \verb|<n>| ranges from 1 to 3,
and \verb|[nprocs]| is an optional argument that indicates the number of
threads to use (default to 1). Full instructions, including those for running
those experiments without Docker, are provided in the \texttt{README}
files.

The compute-hours below are estimated for
$\texttt{nprocs} = 16$.

\bigskip
\begin{compactitem}
    \item[(E1):]
    [Compatibility and Performance] [$2$ compute-hours]
    This experiment runs the test cases bundled with the top 100 popular crates on
    \codename{} and supports (C1) and (C4).
    \begin{asparadesc}
        \item[Preparation:] Complete the setup (Section~\ref{subsec:setup}).
        \item[Execution:] Run \verb|./docker-eval1 [nprocs]|.
        \item[Results:] The results will be printed to the standard output.
        They should be arranged in a table and match the results in
        Table~\ref{tab:compat-stats} approximately. In particular,
        greater than $99\%$ of the test cases should pass.
        The performance results should match the result in Section~\ref{subsec:perf},
        with \verb|Performance Average| lower than $50\%$.
        The generated \verb|performance-hist.pdf| should
        reflect a distribution similar to that in
        Figure~\ref{fig:performance-hist}.
    \end{asparadesc}

    \item[(E2):]
    [Effectiveness in Bug Detection] [$0.5$ compute-hours] 
    This experiment runs PoC programs that cause Rust principle violations
    reported in RustSec. It supports (C2).
    \begin{asparadesc}
        \item[Preparation:] Complete the setup (Section~\ref{subsec:setup}).
        \item[Execution:] Run \verb|./docker-eval2 [nprocs]|.
        \item[Results:] The results will be printed to the standard output.
        They should match Table~\ref{tab:benchmarks-cves}.
    \end{asparadesc}
    
    \item[(E3):]
    [Improvements for Mixed Rust Code] [$0.2$ compute-hours] 
    This experiment runs test cases bundled with Rust crates with
    mixed Rust code (i.e., use of FFI or inline assembly) and supports
    (C3).
    \begin{asparadesc}
        \item[Preparation:] Complete the setup (Section~\ref{subsec:setup}).
        \item[Execution:] Run \verb|./docker-eval3 [nprocs]|.
        \item[Results:] 
        The evaluation scripts print out the results to the standard output.
        For all eight crates, at least one entry should be printed where
        \begin{itemize}
            \item Miri indicates ``unsupported operation'' for some test cases;
            \item For some such test cases,
            \codename{} indicates errors such
            as ``Attempting to borrow from an invalid capability.''
        \end{itemize}
        Note that in three of these cases, the bug was discovered when running
        test cases of a different crate than the crate that contains the bug: \\  \texttt{mozjpeg} through \texttt{jpegli},
        \texttt{sxd-document} through \texttt{cargo-wix}, and \texttt{ring} through \texttt{quinn}.
    \end{asparadesc}
  %






\end{compactitem}
\bigskip



\subsection{Notes on Reusability}



The artifacts can help detect Rust principle violations in mixed Rust code.
To use them to this effect, run the scripts \verb|docker-build| and \verb|docker-run|
on your own code.

With some source code modifications to the QEMU component,
one can adjust the behaviour of the artifacts.
The core files are \verb|op_helper.c| and \verb|cap_rev_tree.c|.
For example, the function \verb|borrow_impl()| in
\verb|op_helper.c| implements
the behaviour of borrowing.


\subsection{Version}
Based on the LaTeX template for Artifact Evaluation V20220926.








\fi

\end{document}